\documentclass[aps,prb,reprint,nofootinbib,superscriptaddress]{revtex4-1}
\usepackage{amsmath,graphics,graphicx,epsfig,amssymb,
xcolor,hyperref,latexsym,float,bm,bbm,times,adjustbox,
verbatim,subfigure,mathtools}

\begin{document}

\title{The nonequilibrium thermal state of a voltage biased Mott insulator}

\author{Arijit Dutta} 

\affiliation{Harish-Chandra Research Institiute,
HBNI, Chhatnag Road, Jhunsi, Prayagraj (Allahabad) 211019, India}
\affiliation{Institut f\"ur Theoretische Physik, 
Goethe-Universit\"at, 60438 Frankfurt am Main, Germany}

\author{Pinaki Majumdar}
\affiliation{Harish-Chandra Research Institiute,
HBNI, Chhatnag Road, Jhunsi, Prayagraj (Allahabad) 211019, India}

\date{\today}

\begin{abstract}
We establish the nonequilibrium thermal phases of a voltage driven 
antiferromagnetic Mott insulator in three dimensions, realised 
at steady state under a voltage bias. Starting from the Keldysh action 
for the half filled Hubbard model we derive an effective Langevin 
equation for the `slow' magnetic variables. The coupling of electrons
to these degrees of freedom determine the transport properties.  At low 
temperature we find a voltage-driven discontinuous insulator-metal 
transition, along with hysteresis. We map the suppression of the N\'eel 
temperature $T_N$ and pseudogap temperature $T_{pg}$ with increasing 
voltage, and discover that the biased Mott insulator has a finite 
temperature insulator-metal transition. The low temperature results 
resolve an experimental puzzle about hysteresis, and the thermal 
results make testable predictions on spectra and nonlinear 
transport.  
\end{abstract}

\maketitle

\section{Introduction}

Strongly correlated systems driven out of equilibrium define a 
frontier in condensed matter. Experiments have probed 
the response to large bias in Mott insulators
\cite{kumai,cro-2013,cro-2019-II,wang,zimmers,radu,
vo2-2011,vo2-2015,magnetite-2007,magnetite-2009,cro-2019}, 
the effect of intense pulsed radiation in `pump-probe' 
experiments \cite{okamoto-nmat-2017,toda,kubler}, and 
metastable hidden phases \cite{stojchevska,xun,ivan,sun,zhang}. 
Among these, the voltage biased Mott insulator is widely studied 
due to the well understood equilibrium state and the 
occurence of a bias driven insulator-metal transition (IMT). 
The breakdown of the `collectively localised' Mott state is 
expected to be very different from that of a band insulator. 

Experiments across multiple materials suggest that the
current-voltage ($I$-$V$) chracteristics in Mott insulators have 
some generic features
\cite{vo2-2011,vo2-2015,cro-2013,cro-2019,
magnetite-2007,magnetite-2009,sabeth}. These are
(i)~a low temperature hysteresis in the current with respect 
to voltage sweep -  changing abruptly from low current to 
high current at some voltage $V_c^+$ on the 
upward sweep, and showing the reverse
switching  at $V_c^- < V_c^+$ on the downward sweep, and
(ii)~reduction of $V_c^{\pm}$ 
and also $\Delta V_{c} = V_c^+-V_c^-$ with
increasing temperature, with hysteresis vanishing 
above some temperature $T^*$.
These features have been observed in samples  of
nanometer \cite{magnetite-2007} to millimeter 
\cite{cro-2019} size.
Scanning near field optical microscopy (s-SNOM) measurements
reveal that the voltage induced breakdown has a progressive
spatial character\cite{cro-2019-III}.

Multiple theories have tried to model the voltage
induced breakdown \cite{aoki-tdschr-2003,
aoki-dmrg-2005,dagotto,okamoto,eckstein1,neqDMFT,amea,diener,
frohlich1,frohlich2,frohlich3,lee,mazza1,li,han,tanaka,
rubtsov,neqmf,schuller,sethna}.  Most microscopic approaches 
suggest a Landau-Zener (LZ) like mechanism
\cite{aoki-tdschr-2003,aoki-dmrg-2005,dagotto,eckstein1,
mazza1}. The resulting $I$-$V$ fails to capture
the discontinuous nature of breakdown, and the strong
temperature dependence observed in a wide variety of 
compounds. Phenomenological network models invoking the
ideas of percolation \cite{schuller,sethna} capture the
low $V$ transport for some materials 
but their applicability in the strongly nonequilibrium state
remains uncertain.
It is only for narrow gap `dirty' Mott insulators, with 
in-gap states, that a successful theory \cite{diener} 
based on ideas of 
Frohlich \cite{frohlich1,frohlich2,frohlich3} seems to
be available. 

The main limitation of current methods 
arise from the neglect of broken translation symmetry (due 
to the bias) and the difficulty in accessing the long time 
steady state.
This paper addresses the metallisation of a Mott insulator,
in a three dimensional (3D) geometry,
using a method that is non 
perturbative in both the interaction strength and the applied 
bias and handles thermal fluctuations exactly. 
Our Keldysh based Langevin dynamics approach exploits
the `slowness' of the magnetic fluctuations on electronic
timescales, retains the effects of dissipation channels (the 
leads) and yields the nonequilibrium electronic state 
at long times. 
The approach is a `twofold' generalisation of the standard
magnetic mean field theory of the Hubbard model: (i)~at 
zero temperature $(T=0)$ we get a Keldysh mean field theory
for magnetism in the biased open system, while (ii)~at finite
temperature a `thermal noise' generates magnetic fluctuations 
in the driven system. The result is a stochastic evolution 
equation for the magnetic moments ${\vec M}_i(t)$ (see
later) which define the background for electron physics.

We work with the half filled Hubbard model in a 3D 
geometry, set the onsite repulsion $U/t=6$, where $t$ is the 
nearest neighbour hopping, and probe the bias $(V)$ and 
temperature $(T)$ dependence of the nonequilibrium state. 
At $V=0$ our approach yields a N\'eel transition at 
$T \sim 0.28t$ which compares well with the quantum Monte
Carlo value of $T_N \sim 0.3t$. 

On introducing the bias we discover the following.
(i)~The $T=0$ state shows a voltage sweep dependent transition 
at $V_c^{\pm}$, between an antiferromagnetic insulator (AF-I) 
and a paramagnetic metal (P-M).  The hysteretic window narrows 
with increasing temperature and vanishes at $T_{coex} \sim 0.02t$.
(ii)~The Neel temperature $T_N$ reduces slowly with
$V$ upto $V \sim 0.5 V^+_c$ and then falls sharply
- vanishing at $V \lesssim V_c^+$. This correlates
with a thermally induced broad distribution of moment 
magnitudes, with a low mean value, at large $V$. 
The pseudogap formation
temperature $T_{pg}$, signalling the crossover from gapped 
to pseudogapped density
of states (DOS),  follows a trend similar to $T_N$.
(iii)~Apart from the expected insulating and metallic
temperature dependence at small and large $V$,
respectively, we observe a thermally driven 
metallisation of the Mott insulator at intermediate 
bias. (iv)~We show that thermally induced amplitude
fluctuation of the moments, and a suppression of mean
magnitude, when $V \rightarrow V_c$, is the primary
driver behind the collapse of $T_N$, $T_{pg}$, and the
thermally induced IMT.

The paper is organised as follows. We start by defining
the model for the open system that describes the 
biased Mott insulator, and introduce the dynamical
equation at the heart of our method. The next
section highlights results on the nonequilibrium
$V-T$ phase diagram, the current response $I(V,T)$,
and the density of states $A(\omega, V, T)$. This 
if followed by a Discussion section, that places 
our method in the context of other approaches to
the correlated electron problem, an analysis of the
magnetic configurations that arise with varying
$V$ and $T$, and the numerical checks that we 
have implemented. A Supplement\cite{supp}
shows how the Langevin equation arises in the
`semiclassical' limit from a Keldysh field theory.

\section{Model and method}

\subsection{Model}

The Hamiltonian for system, baths and their coupling is,
\begin{eqnarray}
\mathcal{H}_{tot} ~~&=& \mathcal{H}_{Hubb} + \mathcal{H}_{bath} + 
\mathcal{H}_{coupl} \cr
\cr
\mathcal{H}_{Hubb} &=& -t\sum_{<ij>,\sigma}
\left(d^{\dagger}_{i\sigma}d_{j\sigma} + h.c.\right)
+ U\sum\limits_{i} n_{i\uparrow}n_{i\downarrow}\cr
\cr
\mathcal{H}_{bath} 
& =& \sum_{\nu,\sigma,\beta\in\{L,R\}}\epsilon_{\nu}
c^{\dagger\beta}_{\nu\sigma}c^{\beta}_{\nu\sigma}\cr
~~\cr
\mathcal{H}_{coup} &=& -\sum\limits_{\substack{<ij>,\sigma}}v_{ij}
\left(c^{\dagger L}_{i\sigma}d_{j\sigma} + c^{\dagger R}_{i\sigma}
d_{j\sigma} + h.c.\right)
\end{eqnarray}
In the equations above
$n_{i\sigma}=d_{i\sigma}^{\dagger}d_{i\sigma}$ and 
$\epsilon_{\nu}$ are the bath eigenenergies. $v_{ij}$ denote 
the system-bath couplings. We assume the density of states of 
the bath to be a Lorentzian. The chemical potential $\mu$ in 
the system is tuned to ensure half-filling. A voltage bias is 
applied by tuning the chemical potential in the left (right) 
leads $\mu_{L(R)}$. We set 
$\mu_{L(R)} = \mu\pm\left(V/2\right)$.

\subsection{Method}

Starting from the Keldysh\cite{kamenev} action for the above 
Hamiltonian (see Supplement\cite{supp}) 
we decouple the quartic term by 
Hubbard-Stratonovich transformation. This introduces 
real auxiliary fields at each instant,  
henceforth called the charge field $\phi_i(\tau)$ and 
spin field $\vec{M}_i(\tau)$, that 
couple respectively to the instantaneous 
density and spin of the electrons.
$\tau$ is our time variable, $t$ being
used for the hopping. The action becomes quadratic in the 
Grassmann fields which can be formally integrated out. 
We fix $\phi_i$ to its half-filling equilibrium 
saddle point value. Using assumptions related to the slowness 
of the ${\vec M}_i(\tau)$, and a simplified noise kernel, both 
discussed in the Supplement\cite{supp}, we derive a 
stochastic dynamical equation for ${\vec M}_i(\tau)$. 
\begin{eqnarray}
{{d\vec{M}_i} \over {d \tau}} ~-~ \alpha
  &&\left(\vec{M}_{i} \times\frac{d\vec{M}_{i}}{d\,\tau}\right)
 = \gamma_{i}(\tau)\left(\langle {\vec \sigma}_i
 \rangle_{\{\vec M\}} - \vec{M}_{i}+  \vec{\xi}_i\right) \cr
\cr
&&\langle\xi^{a}_{i}(\tau)\xi^{b}_{j}(\tau^{\prime})\rangle 
= \frac{4 T}{U\gamma_i(\tau)}\delta_{ij}\delta_{ab} 
\delta(\tau - \tau^{\prime})  \cr
\cr
&&\langle  {\vec \sigma}_i \rangle_{\{\vec M(\tau)\}} 
= \int\mathrm{d}\omega\,\mbox{Tr}\left[
\hat{\mathcal{G}}^{K}_{ii}(\tau,\omega)
\vec{\sigma_P}\right]
\end{eqnarray}
where $a,b$ denote $O(3)$ indices.
From consistency arguments, discussed in 
the Supplement\cite{supp}, 
$\gamma_{i}(\tau) = \frac{2U}{\alpha}(1+\alpha^2|
\vec{M}_{i}(\tau)|^{2})$, where $\alpha = (U/t)^{2}$ is 
the \emph{Gilbert damping}\cite{gilbert,lakshmanan}. 

$\hat{\mathcal{G}}^{K}$ denotes 
the \emph{adiabatic} Keldysh Green's function, and the trace 
is over the local $2\times 2$ spin subspace (assumed 
henceforth). ${\vec \sigma}_i = 
\frac{1}{2}\sum\limits_{\alpha\beta}d^{\dagger}_{i,\alpha}
\vec{\tau}_{\alpha\beta}d_{i,\beta}$,
$\vec \sigma_P \equiv (\sigma^{x},\sigma^{y},\sigma^{z})$ being 
the $2\times 2$ Pauli vector, is the local fermion spin. Its 
average is computed on the instantaneous $\{\vec{M}\}$ 
background. $\langle \vec{\sigma}_{i} \rangle$ is a 
nonlinear, non local, function of the
$\vec{M}$ field and encodes the strong correlation effects
in the problem. The computationally hard part in the evolution equation is 
calculation of 
$\hat{\mathcal{G}}^{K}_{ii}(\tau,\omega)$, briefly indicated
next. We also the need 
$\hat{\mathcal{G}}^{<}_{i+\hat{x},\,i;\sigma}(\tau,\omega)$ for
the transport calculation, and
$\hat{\mathcal{G}}^{R}_{ii}(\tau,\omega)$ for the
density of states.
\begin{align}
	\hat{\mathcal{G}}^{R}(\tau,\omega) &= \left(\omega - 
	\hat{\mathcal{H}}(\tau) + \hat{\Gamma}^{R}(\omega)\right)^{-1}
	=\left[\hat{\mathcal{G}}^{A}(\tau,\omega)\right]^{\dagger}\nonumber\\
	\hat{\mathcal{H}}_{ij}(\tau) &=\,(-\vec{M}_i(\tau)\cdot\vec{\sigma}^P+\phi_i)
	\delta_{ij}-t_{<ij>}\nonumber\\
	\left[\hat{\mathcal{G}}^{-1}\right]^{K}_{ij;\alpha\beta}(\tau,\omega) &=
	-2i|v|^2\rho_{B}(\omega)\Biggl[\tanh\left(\frac{\omega-\mu_L}{2T}\right)
	\delta_{i,i_L}\nonumber\\
	&~~~~~~~~~~+\tanh\left(\frac{\omega-\mu_R}{2T}\right)\delta_{i,i_R}\Biggr]
	\delta_{ij}\delta_{\alpha\beta}\nonumber\\
	\hat{\mathcal{G}}^{K}(\tau,\omega) &= \hat{\mathcal{G}}^R(\tau,\omega)
	\left[\hat{\mathcal{G}}^{-1}\right]^K(\tau,\omega)
	\hat{\mathcal{G}}^A(\tau,\omega)\nonumber\\
	\hat{\mathcal{G}}^{<}(\tau,\omega) &= \frac{1}{2}\left[
	\hat{\mathcal{G}}^K(\tau,\omega)-\hat{\mathcal{G}}^R(\tau,\omega)
	+\hat{\mathcal{G}}^A(\tau,\omega)\right]
\end{align}
where $\rho_B(\omega)$ is the density of states of the baths,
$i_{L(R)}$ denote the sites at the left (right) edge.

The Langevin equation is solved  using a stochastic
Heun discretisation scheme \cite{palacios} to generate a 
time series for ${\vec M}_i(\tau)$. Upon obtaining the time 
series the electronic observables are
computed on the instantaneous configurations (assuming that
electronic timescales are much shorter than spin fluctuation 
scales) and averaged over the time series.

We benchmark the scheme against equilibrium 
Monte-Carlo results on the adiabatic problem
(see Appendix~\ref{appendixB}). 
The nonequilibrium results pertain to a $8\times 4\times 4$ 
system, with $L=8$ being the longitudinal (transport) 
direction. We discuss size dependence in the Discussion section.
Starting with an arbitrary $\{\vec{M}\}$ 
configuration the system is evolved for $\sim 10^{6}$
steps with a time discretisation of $10^{-3}\tau_{0}$, where
$\tau_{0} \sim 1/J_{eff}$ (with $J_{eff}\sim t^{2}/U$) is the
characteristic timescale of the auxiliary field. After 
allowing the system to equlibriate for $100\tau_{0}$ 
(this is taken to be $\tau=0$ in the
definition of time averaging of observables), 
the rest of 
the configurations have been saved to 
construct the time series for ${\vec M}_i(\tau)$. 
The maximum time of the simulation is 
$\tau_M\sim 1000\tau_0$.
\begin{figure}[t]
\centerline{
  \includegraphics[width=4.5cm,height=4.7cm]{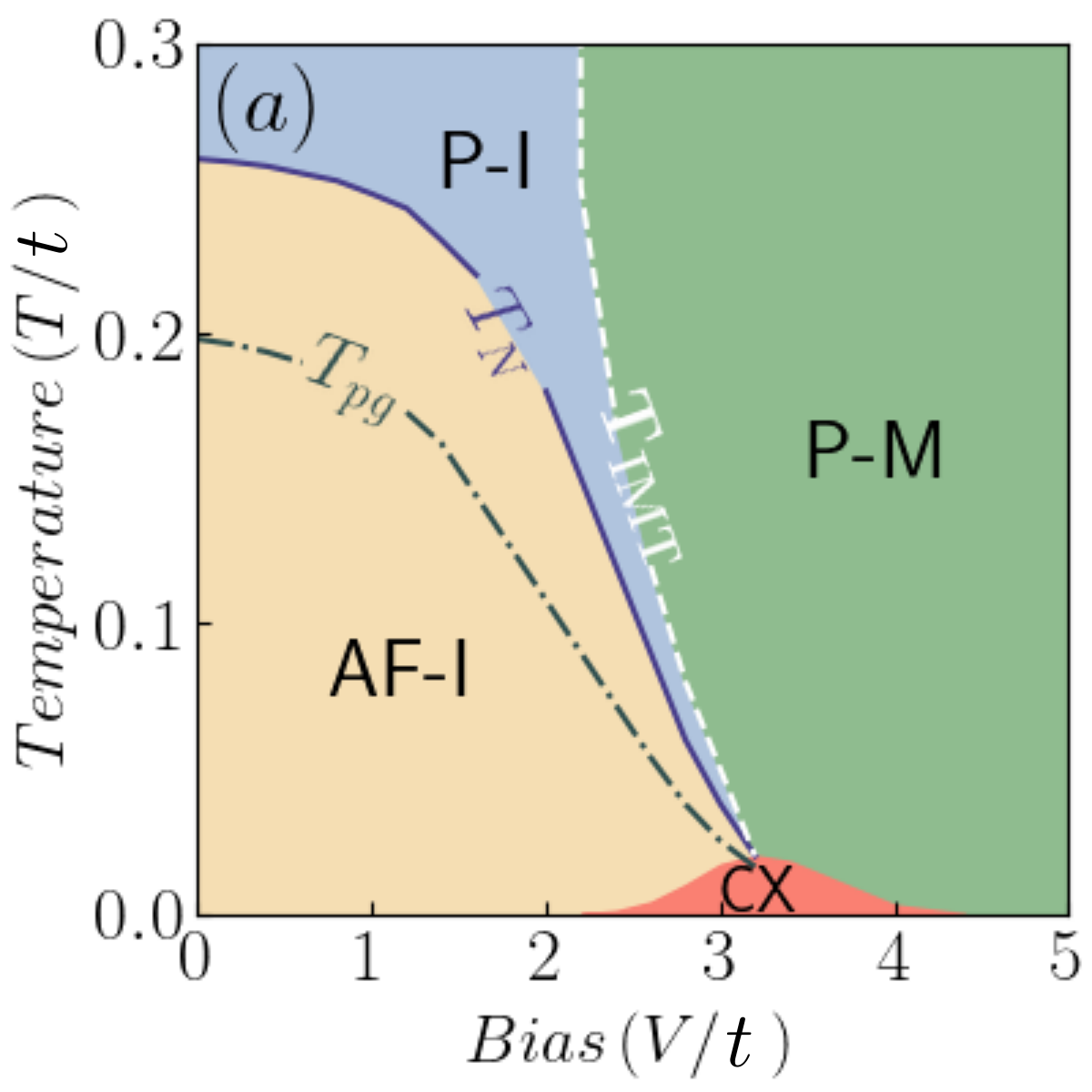}
  \includegraphics[width=4.5cm,height=4.7cm]{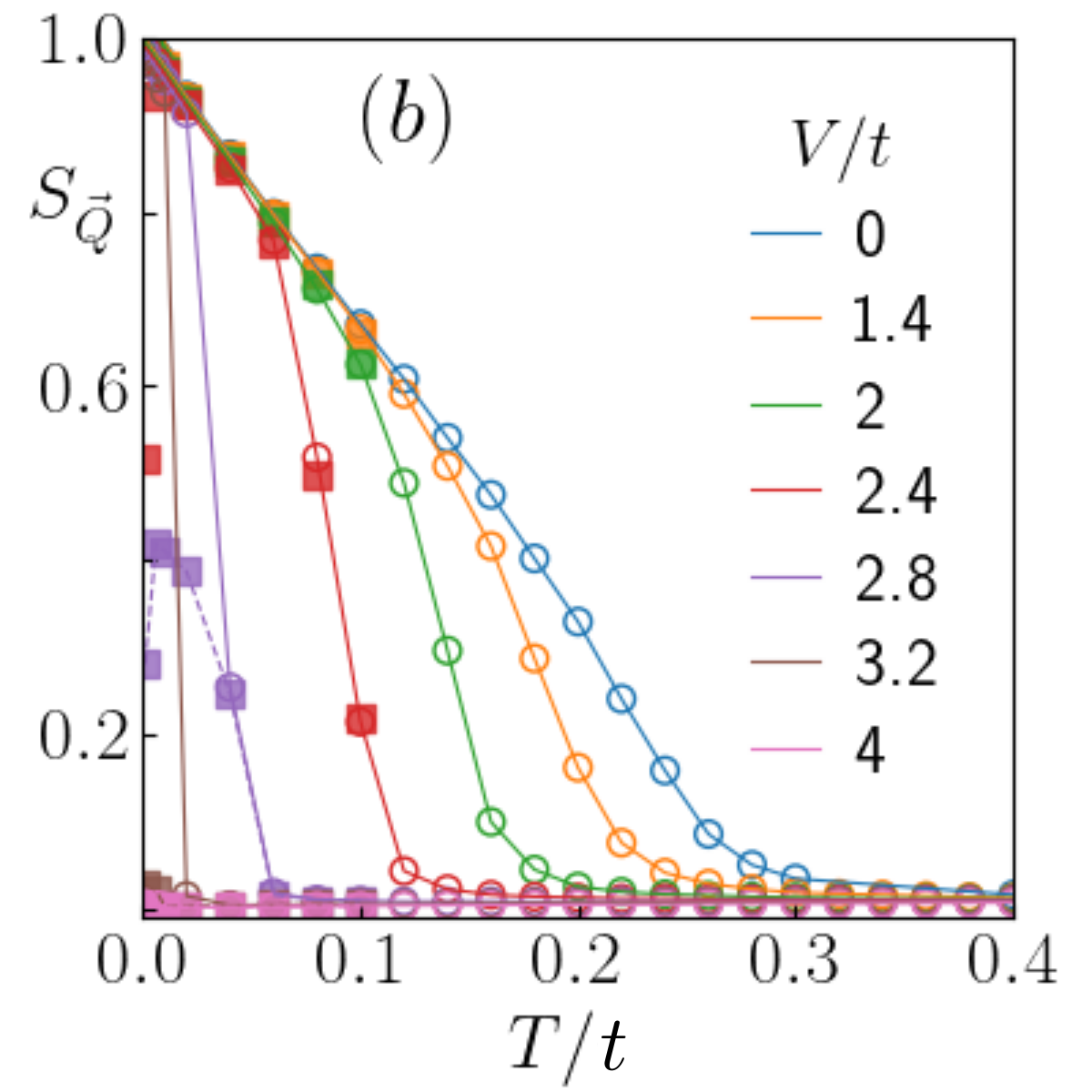}
}
\caption{(a)~Temperature ($T$) vs voltage ($V$) phase diagram
of the voltage biased repulsive Hubbard model at $U/t = 6$. 
The AF-I, P-M and P-I are the antiferromagnetic insulator, 
parmagnetic metal and paramagnetic insulator phases 
respectively. Insulating (metallic) regimes have 
${\partial I}/{\partial T} >(<) 0$, where $I$ is the 
steady state current. CX marks the hysteretic window. 
The solid blue line indicates $T_{N}(V)$, the dashed white 
line $T_{IMT}(V)$ and the broken grey line indicates 
$T_{pg}(V)$. (b)~The magnetic ordering peak, 
$S_{\vec{Q}}(T)$ for upward (open circles) and downward 
(solid squares) $V$ sweeps. For $T/t \geq 0.02$ the 
two curves coincide for all values of $V$. The
inflection point for each curve gives the $T_N$ for the 
corresponding $V$.}\label{fig1}
\end{figure}

\section{Results}

\subsection{Phase diagram}

The $V=0$ ground state at half-filling is an 
antiferromagnetic insulator (AF-I) for 
any finite $U$ \cite{qmc,anamitra}. $|\vec{M}|$ 
grows with increasing $U$ and saturates to unity as $U/t 
\rightarrow \infty$. As the temperature is increased the
system  loses long range order (LRO) at a scale $T_N(U)$. 
For $T > T_N$, the system
is a paramagnetic metal (P-M) for  $U/t \lesssim 4$ and 
a paramagnetic insulator (P-I) for $U/t \gtrsim 4$.
The crossover region from P-M to P-I shows a 
pseudogapped density of states (DOS).
The equilibrium physics and how our method accurately 
captures
it is discussed in detail in the Discussion section.

We construct a nonequilibrium $V-T$ phase diagram
at $U=6t$, Fig.\ref{fig1}(a),  highlighting the magnetic, 
transport and spectral regimes that occur in the biased 
problem.  There are three phases, AF-I, P-I and P-M, and a 
low $T$ coexistence (CX) window bounded by $V^{\pm}_c$.
The bias dependent temperature scales are $T_N$, for the
magnetic transition, $T_{IMT}$ for the narrow window of
thermally induced insulator-metal transition, and $T_{pg}$
for crossover from gapped to pseudogap DOS.
The indicators in terms of which we infer magnetic order,
transport behaviour, and spectral features, are
discussed below.

Fig.\ref{fig1}(b) shows the peak, $S_{\vec Q}(T)$ in the 
magnetic structure factor $S_{\vec{q}}$, where $S_{\vec{q}}$ 
is defined by,
\begin{equation}
S_{\vec{q}} = 
{1 \over N^2} \sum_{ij} 
\int\limits_{0}^{\tau_{M}}\frac{\mathrm{d} \tau}{\tau_{M}} 
\vec{M}(\tau,\vec{r}_{i}) \cdot \vec{M}(\tau,\vec{r}_{j})
e^{i \vec{q}\cdot(\vec{r}_{i}-\vec{r}_{j})}
\end{equation}
The $\tau=0$ point corresponds to the start of the
measurement period in the nonequilibrium steady state
and is $100\tau_0$. 
$\vec{q} = \vec{Q}= (\pi,\pi,\pi)$ pertains to 
N\'eel AF order. For the cubic lattice with nearest
neighbour hopping the peak in the
structure factor remains at $(\pi,\pi,\pi)$
at all $V$.
For each $V$ the N\'eel temperature is 
estimated from the point of inflection of the 
$S_{\vec Q}(T)$ curve. There exists a coexistence region at 
$T=0$ for $2.2 \leq V/t < 4.3$,
which extends upto $T_{coex}/t = 0.02$. The state in this 
region depends on the direction of voltage sweep. 
In Fig\ref{fig1}.(b), the open circles denote
the $T$ dependence of $S_{\vec{Q}}$ for upward sweep, 
while the solid squares denote the same for the downward 
sweep. The solid blue curve in Fig.\ref{fig1}(a) shows the 
dependence of $T_N$ on $V$. $T_N$ decreases slowly 
initially, with $V$, and then quicker for 
$V \gtrsim 0.5 V^+_c$ and 
vanishes for the upward sweep at $V/t \approx 3.6$.

\begin{figure}[b]
\centering
\includegraphics[width=8cm,height=9cm]{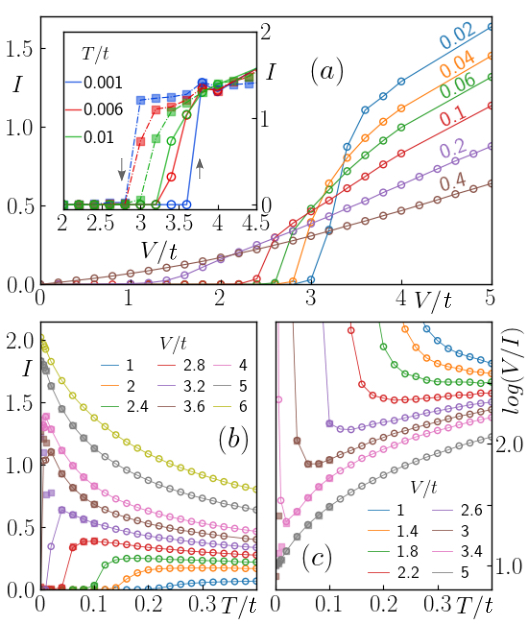}
\caption{(a) The current-voltage ($I$-$V$) characteristics 
with changing temperature ($T$). The hysteretic behaviour 
at low $T$ is shown in inset. The solid (dashed) 
lines and open circles (filled squares) correspond to 
upward (downward) voltage sweeps. The arrows indicate the 
sweep direction. For $T < 0.02t$, $I$ 
changes discontinuously at $V_{c}^{\pm}(T)$ for the upward 
and downward sweeps, respectively. Beyond the coexistence 
region the $I$-$V$ has a unique threshold at $V_{c}(T)$ 
which reduces with increasing $T$ and vanishes for 
$T > 0.1t$. (b) $I$ vs $T$ for different $V$.
$\partial I/\partial T >(<)\,0$ indicates an insulating 
(metallic) phase. A peak in the $I(T,V)$ curve for a fixed 
$V$ indicates a temperature driven IMT. (c) Log of resistance 
($R = V/I$) vs $T$ for different $V$. Analogously, a minimum 
in the $R(T,V)$ curve for a fixed $V$ indicates IMT.}
\label{fig2}
\end{figure}
\begin{figure*}[t]
\centerline{
~~~
\includegraphics[width=15.3cm,height=5.8cm]{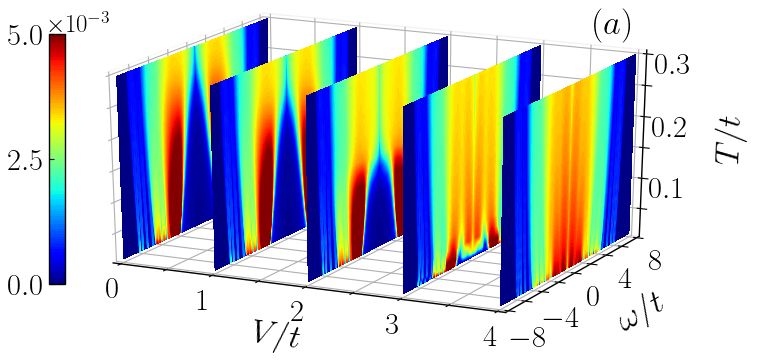}
}
\vspace{.4cm}
\centerline{
\includegraphics[width=14.3cm,height=5.0cm]{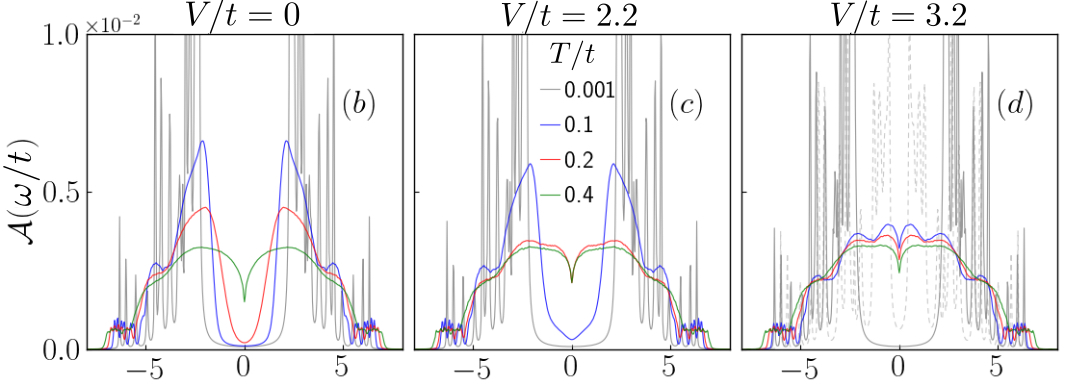}
}
\caption{(a) Map of DOS for varying temperature
and voltage, on the upward sweep. (b-d) Temperature
variation for $V/t = 0, 2.2$ and $3.2$. For $V \leq 2.2t$,
the low $T$ DOS remains gapped and becomes pseudogapped
for $T > T_{pg}$, in both the sweep cycles.
For $2.2 < V/t < 3.4$ the DOS remains gapped at low
temperature, develops subgap weight with increasing
$T$, even in the insulating phase, and becomes pseudogapped
at large $T$. For the downward sweep, the DOS in this regime
retains subgap weight even to the lowest temperature,
as shown in (d) with a dashed line. For $V \geq 3.4t$ the DOS
remains ungapped at low $T$ and broadens with increasing $T$.
}\label{fig3}
\end{figure*}

\subsection{Transport}

The charge current across a transverse cross-section at site 
$j$ is given by
\begin{eqnarray}
I_{x}\left(V\right)~~ &=&~ 
\sum_{y,z}
\int\limits_{0}^{\tau_M}
\frac{\mathrm{d} \tau}{\tau_M}\int\limits_{-D}^{D}
\frac{\mathrm{d}\,\omega}{2\pi}\,
\mbox{Tr} [\delta \mathcal{G}^{<}_{j+\hat{x},\,j;\sigma}(\tau,\omega)] \cr
\delta \mathcal{G}^{<}_{j+\hat{x},\,j;\sigma} & = &
~~ \mathcal{G}^{<}_{j+\hat{x},\,j;\sigma}(\tau,\omega)~ 
- ~\mathcal{G}^{<}_{j,\,j+\hat{x};\sigma}(\tau,\omega)
\end{eqnarray}
where $\mathcal{G}^{<}$ is the adiabatic \textit{lesser} 
Green's function and the sum is over all sites in the 
transverse cross-section containing the site $j$. Due to 
charge conservation $I_{x}(V)$ must be independent of $x$.
However, there is a weak violation ($< 10\%$) 
of current conservation at very low temperatures due to a finite
convergence factor $\eta\leq 0.01t$, which is needed for
numerical stability of the scheme. The current conservation 
is better satisfied with increasing $T$ and $V$.
The $I$-$V$ characteristics are 
plotted in Fig.\ref{fig2}(a) for different temperatures. The 
inset shows hysteresis for $T < T_{coex} \sim 0.02 t$ while 
the main panel shows the response for $T/t \gtrsim 0.02$. 
Above $T_{coex}$ and upto $T/t \sim 0.3$ it has a `threshold'
at some $V_{c}(T)$ (below which the current remains 
exponentially suppressed) that reduces with increasing $T$. 
Beyond $V_{c}$, the current rises sharply with increasing 
$V$ and saturates as $V$ approaches the bandwidth $D$ of the 
connected system. The current saturation at large $V$ is 
similar to what has been observed in the 2D problem 
at zero $T$ \cite{neqmf}. The suppression of 
$V_c$ with increasing $T$ has been observed in 
experiments on various driven Mott systems
\cite{kumai,vo2-2011,vo2-2015,magnetite-2007,magnetite-2009,
cro-2013,cro-2019}.

Figs.\ref{fig2}(b) show $I(T)$ at different $V$ .
The results reveal three regimes:
(i)~insulating, where the system becomes more
conducting with increasing $T$, i.e., 
$\partial I/\partial T > 0$
at all $T$ (happens for $V/t \leq 2$), 
(ii) metallic, showing  $\partial I/\partial T < 0$ at all
$T$ (occurs for for $V/t > 3.8$), and (iii)~showing 
insulator to metal transition: $\partial I/\partial T$ 
changing sign at $T_{IMT}$. This happens for $2 < V/t < 3.8$.
The corresponding `resistance' $R = V/I$ is shown in 
Fig.\ref{fig2}(c) on a
logarithmic scale. In the deep insulating regime $R$ 
decreases exponentially with increasing $T$ and in the strongly
metallic regime it rises monotonically with $T$. At intermediate $V$ 
it shows non monotonic $T$ dependence. This feature, arising 
from thermal fluctuations in a non equilibrium situation, 
is the most important result of our paper. We will discuss
the physical basis further on. Note that within a linear 
response treatment of the Mott insulator $V/I$ is 
independent of $V$ and solely dependent on $T$. This
would be true of the $V/t \lesssim 1$ window (the top right 
curve). The effective resistance at all other voltages 
depends crucially on the applied bias.

\begin{figure*}
\centerline{
\includegraphics[width=14cm,height=7cm]{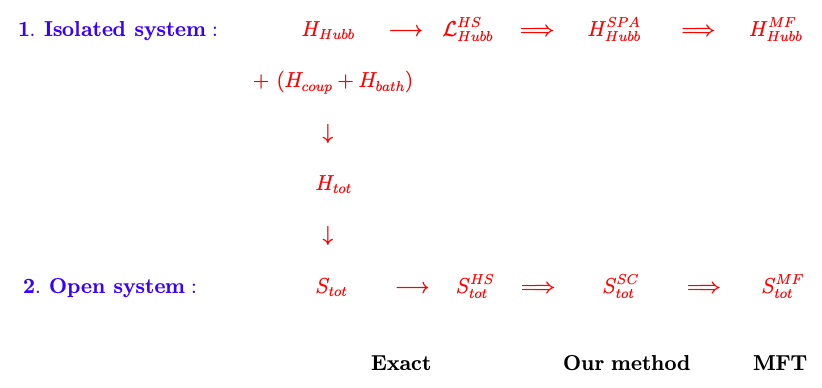}
}
\caption{Sequence of approximations in the isolated equilibrium
system (top)  and the open nonequilibrium system (bottom).
The single arrows indicate exact transformations while the
double arrows indicate approximations. Our method, described
in the text, approximates the magnetic variables as ``slow''
but retains fluctuation effects far beyond mean field theory (MFT).
The notation is described in the text below.
}
\end{figure*}

\subsection{Density of states}

The system averaged single particle density of states (DOS)
is given by 
\begin{equation}\label{eq:dos}
\mathcal{A}(\omega)=-\frac{1}{2\pi N}\sum_{i}\int
\limits_{0}^{\tau_{M}}\frac{\mathrm{d} \tau}{\tau_{M}}\,\mbox{Im}
(\mbox{Tr}[\mathcal{G}^{R}_{ii}(\tau,\omega)])
\end{equation}
where $\mathcal{G}^{R}$ is the adiabatic \textit{retarded} 
Green's function. $N$ is the total no. of sites. Its 
behaviour with increasing $T$ in different voltage regimes 
is shown in Fig.\ref{fig3}(a). For $0 \leq V/t \leq 2.2$
(Fig.\ref{fig3}(b),(c)), at low $T$, the DOS has a gap 
independent of the sweep direction, which gets smeared 
with increasing $T$ and ultimately becomes a pseudogap 
beyond $T_{pg}$. For $2.2 < V/t < 3.4$ (Fig.\ref{fig3}(d)), 
in the upward sweep the DOS remains gapped at low 
temperatures, but develops subgap weight upon 
increasing $T$. Upon heating beyond 
$T_{IMT}(V)$ the DOS becomes pseudogapped and
 broadens with 
increasing $T$ further. For the downward voltage sweep in 
this regime, the DOS remains ungapped even to the lowest 
temperature, as shown in Fig.\ref{fig3}(d) with a dashed 
line.

\section{Discussion}

In our Model and Method section we introduced the Langevin 
approach without any discussion of its place within the larger 
scheme of many body theory for 
nonequilibrium systems. Similarly, most of the results presented
till now has been numerical data, without much analysis of why we
observe a certain kind of behaviour. 
We adopted this approach to get across the 
basic results quickly without digression.
This section, the Appendices, and the Supplement\cite{supp}, aim to
fill up the gaps, by placing our method in context, 
and motivating the results we have shown.

The first subsection focuses on the method that we have used.
It addresses where our method lies in the spectrum between 
full fledged quantum Monte Carlo (QMC) and simple mean field
theory (MFT), and the benchmarks it satisfies at equilibrium.

The subsections thereafter focus on features of the magnetic 
configurations
${\vec M}_i(\tau)$ that arise from the Langevin evolution.
Since the electron response time is assumed
to be much shorter than the magnetic fluctuation time the 
electronic properties are computed on magnetic configurations
on individual `time slices', $\tau$, and then averaged.

In what follows we (i)~locate our method within the larger
family of many body approximations and distinguish it from 
mean field theory, (ii)~highlight some aspects of the distribution 
$ P(M,V,T)$,  (iii)~suggest a Landau like functional at $T=0$ for 
the bias driven first order transition, (iv)~propose a plausible
mechanism for the suppression of $T_N$ with $V$, and (v)~quantify
the scattering mechanism that seems to decide the current in 
the bias stabilised metallic phase. 
We also comment (vi)~on the spatial variation of the
electron density and local moment magnitude and (vii)~on 
the size and dimension dependence of our results. Finally, 
(viii)~we briefly discuss the connection of our results to 
experiments on bias driven Mott materials.

\vspace{-.1cm}

\subsection{Locating our method}

In our methods section we directly moved to the Langevin
equation in the nonequilibrium problem. This would be 
a case of double unfamiliarity for many readers since
(i)~the Langevin approach is not standard even in the
equilibrium problem, and (ii)~the driven problem 
additionally complicates the formulation by bringing
in leads and a bias.  In what follows we quickly discuss
the equilibrium formulation that generalises to our
nonequilibrium scheme, differentiate it from simple
mean field theory, and schematically show how the
nonequilibrium method arises. The detailed derivation
of the nonequilibrium scheme is given in the Supplement\cite{supp}.

\subsubsection{At equilibrium}

The equilibrium problem corresponds to disconnecting
the Hubbard block from the leads (and setting the bias $V=0$). 
There are three levels at which the Hubbard model can
be approached. The flow chart in Fig.4 illustrates these.
$$
H_{Hubb} 
~\rightarrow 
~{\cal L}_{Hubb}^{HS}
~\rightarrow 
~H_{Hubb}^{SPA}
~\rightarrow 
~H_{Hubb}^{MF}
$$
In the above $H_{Hubb}$ is the Hubbard model, 
${\cal L}_{Hubb}^{HS}$ is the corresponding Lagrangian 
after
rewriting the interaction in terms of charge and spin 
auxiliary fields, $\phi_i(\tau)$ and ${\vec M}_i(\tau)$, 
respectively. Till this is exact. The approximation we
use is to treat the auxiliary fields as only spatially
fluctuating, neglecting the $\tau$ dependence. This is
called the `static path approximation' and leads to 
$H_{Hubb}^{SPA}$. It retains all spatial 
thermal fluctuations but no temporal fluctuations.
A possible further simplification is to drop the
spatial fluctuations as well, retaining only 
the `order parameter mode'. This is the mean field model,
$H_{Hubb}^{MF}$,
retaining only the ${\vec q} = (\pi, \pi, \pi)$ mode
of the ${\vec M}_i$ field. 
A fourth approach, not listed above, corresponds to
dynamical mean field theory (DMFT) which would
retain the temporal fluctuations of the auxiliary
fields and drop the spatial dependence.
We write the exact, the SPA, and the mean field
models below. For SPA in the half-filling
case we ignore fluctuations of the charge field
$\phi_i$.

The models are, successively,
\begin{eqnarray}
H_{Hubb} & = &
-t\sum_{<ij>,\sigma} d^{\dagger}_{i\sigma}d_{j\sigma}
+ U \sum_i n_{i \uparrow} n_{i \downarrow}
\cr
H_{Hubb}^{SPA} &=& 
-t\sum_{<ij>,\sigma} d^{\dagger}_{i\sigma}d_{j\sigma}
+ U \sum_i {\vec M}_i.{\vec \sigma}_i 
+ U \sum_i M_i^2
\cr
H_{Hubb}^{MF} &=&  
-t\sum_{<ij>,\sigma} d^{\dagger}_{i\sigma}d_{j\sigma}
+ U\sum\limits_{i} Me^{i {\vec Q}.{\vec r}_i} \sigma_{iz}
+ U NM^2
\nonumber
\end{eqnarray}
The full model can be studied via exact diagonalisation, 
severely size limited, or determinantal QMC in terms of
the auxiliary fields $\phi_i(\tau)$ and ${\vec M}_i(\tau)$.
The SPA model can be studied by Monte Carlo sampling of
the field ${\vec M}_i$ (the $\phi_i$ being dropped at
half-filling), or by the Langevin approach - assuming
the $M_i$ dynamics to be much slower than electron
dynamics. 
The MF model, as well known, can be immediately diagonalised
due to the assumed periodic nature of the ${\vec M}_i$
background. Mean field theory restricts  ${\vec M}_i$
to $Me^{i {\vec Q}.{\vec r}_i}$, with ${\vec Q} 
= (\pi, \pi, \pi)$, for the Neel state.
 
Within MFT there is only a site independent magnitude to be
determined - the size $M$ of the magnetic moment. There are 
no angular variables anymore. As a result, MFT has only two 
possible phases: (a)~an AF-I when $M \neq 0$, and (b)~a
trivial P-M when $M=0$. Magnetic order and insulating gap
are intimately connected. 
$M$ would vanish when the temperature is comparable
to the $T=0$ gap, for $U \gg t$ this is $T_N^{MF} \sim U$. 
At our parameter point it is $T_N^{MF} \sim t$.
The mean field spectrum is either gapped or tight binding,
there is no pseudogap phase.

Our approach is SPA, when using Monte Carlo, or the 
equivalent Langevin scheme when using dynamics (see
Appendices).
Within these the magnitude as well as the direction 
associated with the ${\vec M}_i$ can fluctuate and 
the electronic properties are computed in these
backgrounds. Retaining these thermal fluctuations leads 
to two major differences in our results compared to mean 
field theory: (i)~The loss of AF order arises from 
angular fluctuations of the moments rather than
$\vert {\vec M}_i \vert \rightarrow 0$. 
The $T_N^{SPA}$ scale we obtain 
compares very well with full
QMC, our result at $U=6t$ is 
$T_N \sim 0.28t$ compared to the QMC value  $\sim 0.3t$.
(ii)~The gap at large $U$ 
is related to the {\it magnitude} of ${\vec M}_i$
not its long range order. The difference between the
$T < T_N$ and $T > T_N$ phases is mainly in
the angular correlation between the moments, not their
magnitude. As a result, even in the $T > T_N$ phase
a gap can survive - this is the paramagnetic
insulator (P-I). When $U/t$ is in the intermediate 
coupling window, as in our case, the $T \gtrsim T_N$ 
window shows a pesudogap rather than a clean gap.

All these features are visible in the $V=0$ results in 
various figures. Fig.1(a) shows the $T_N$ (see Fig.4 of
a QMC reference \cite{staudt}) and also the temperature
where the low temperature gapped phase transits to
a higher $T$ pseudogap phase.
Fig.2(b) shows that at low $V$ insulating behaviour
persists way beyond $T_N$ - the signature of a
P-I phase.  Fig.3(b) shows the
DOS at $V=0$. The plot corresponding to $T=0.4t$
shows the PG in the spectrum. 
These $V=0$ results gave us the confidence to
treat the bias problem using the Langevin scheme.

\subsubsection{Out of equilibrium}

For the nonequilibrium problem $H_{Hubb}$ is augmented
by $H_{coup} + H_{bath}$ as in Eqn.1. There is no longer
a `Hamiltonian' that describes the system degrees of freedom
and we need to use an action. Following the notation of 
Eqn.1 we have the sequence of approximations:
$$
H_{tot} 
~
\rightarrow 
~
S_{tot} 
~
\rightarrow 
~
S_{tot}^{HS}
~
\rightarrow 
~
S_{tot}^{SC}
~
\rightarrow 
~
S_{tot}^{MF}
$$
The sequence from $H_{tot} \rightarrow .. S_{tot}^{SC}$ is
described in the Supplement\cite{supp}. Upto $S_{tot}^{HS}$ the
formulation is exact. Beyond this we do a semiclassical (SC)
expansion, assuming that the ${\vec M}_i$ fields are
much slower than the electrons, to obtain $S_{tot}^{SC}$.
This is the equivalent of the SPA approximation in the
equilibrium problem. One can further simplify the SC
scheme by neglecting dynamics and thermal 
fluctuation altogether to obtain
the mean field action $S_{tot}^{MF}$. This is what we
used at $T=0$ in an earlier paper \cite{neqmf}.

\vspace{.4cm}

\subsection{Moment magnitude distribution}
\label{sec:pm}

At steady state the magnetic configurations are characterised
by their local distribution $P_i(M)$ and the system averaged
distribution $P(M)$, where
\begin{eqnarray}\label{eq:Pm}
P_i(M) &=& {1 \over {\tau_M}}
\int_0^{\tau_M}
d \tau \delta(M - \vert {\vec M}_i(\tau)\vert)
\cr
&&
\cr
P(M) &=& {1 \over N} \sum_i P_i(M)
\end{eqnarray}
The magnitude $ \vert {\vec M}_i\vert$
plays an important role since $\langle M \rangle = \langle \vert
{\vec \sigma} \vert \rangle$ and $M \rightarrow 1$
indicates that the double occupancy $ d_i = \langle n_{i \uparrow}
n_{i \downarrow}\rangle \rightarrow 0$, {\it i.e}, no charge fluctuation,
while $M \rightarrow 0$ implies a tight binding metal.
The $V$ and $T$ dependence  of $P(M)$ is the primary determinant of
insulating and metallic behaviour.

\begin{figure}[t]
 \centering
 \includegraphics[width=\linewidth]{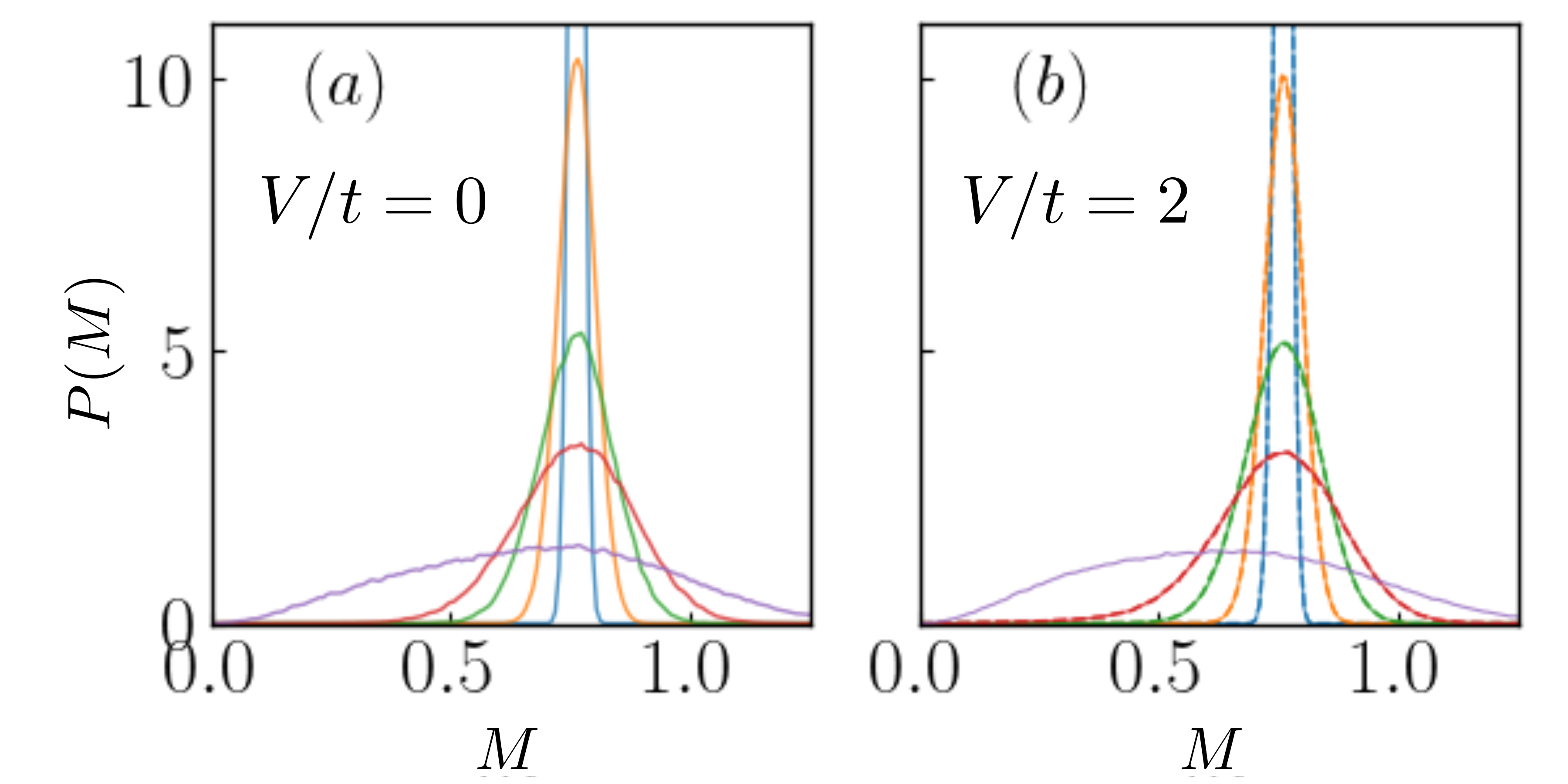}
 \includegraphics[width=\linewidth]{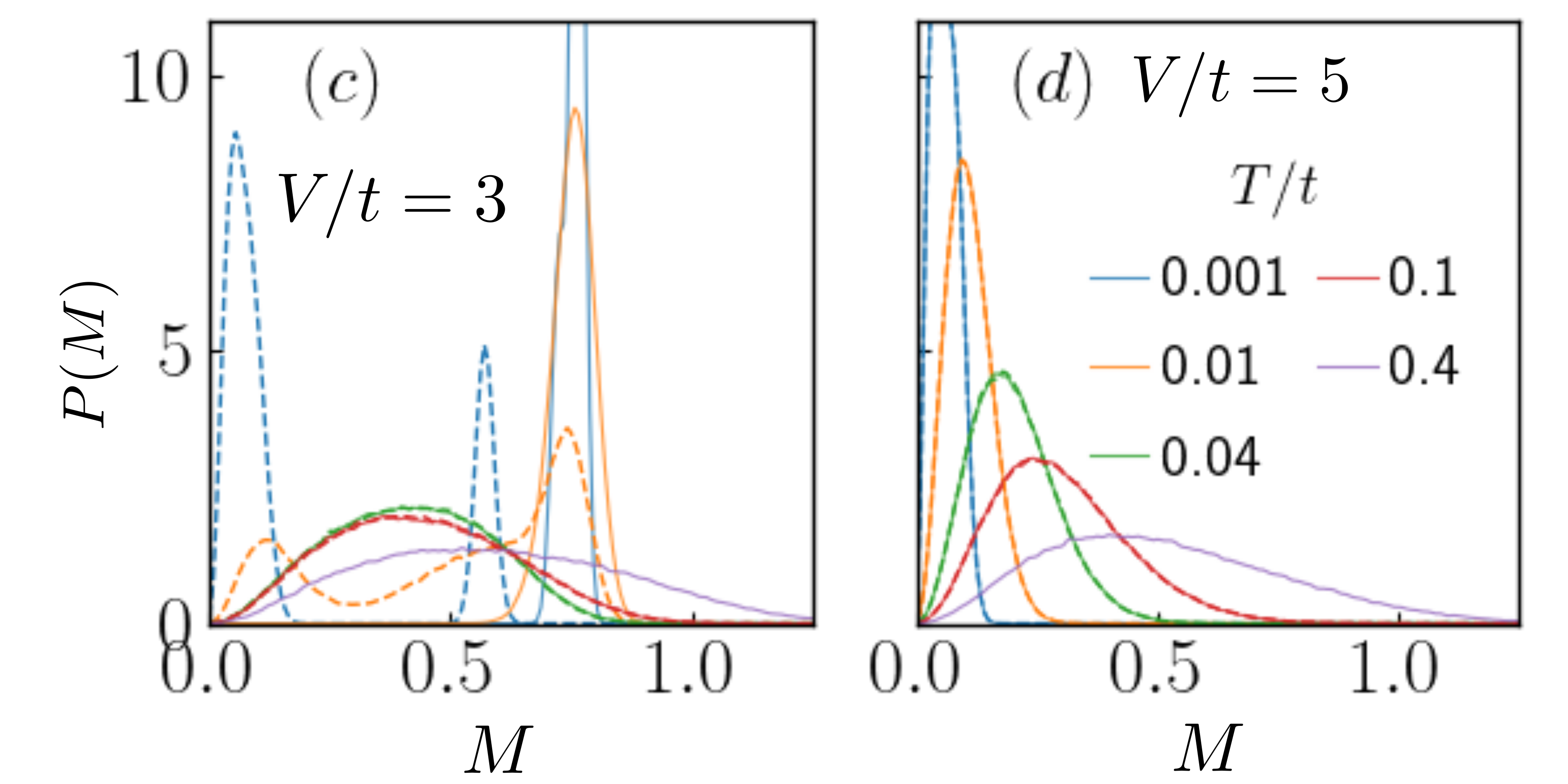}
 \caption{(a-d) Variation of the moment distribution with temperature ($T$)
  for $V/t = 0, 2, 3$ and $5$ respectively. The solid (dashed) lines
  denote the distribution for upward (downward) sweeps at different
  temperatures.
}
\label{fig:pm}
\end{figure}

The effect of bias and temperature on $P(M)$ 
has been plotted in Fig.\ref{fig:pm}. There are
broadly three regimes. (a)~For $V/t \leq 2$, and $T \ll 
T_{N}$, the moments remain almost pinned to their
equilibrium ground state value. Hence, 
the $P(M)$ is sharply peaked around $0.8$.
As temperature increases the distribution broadens around 
the mean value up to $T \approx T_{N}$. Beyond $T_{N}$ the 
distribution becomes skewed and the mean starts shifting to
 lower values with increasing $T$. (b)~Between $2.2 < V/t \leq 4.3$ 
 the low $T$ moment distribution changes from
being unimodal to bimodal, and acquires a sweep dependence. 
This is because the ``effective potential'' governing the moment 
distribution develops a metastable low moment minimum, as we
 shall discuss in the next section. (c)~Beyond $V = 4.3t$,
where the system is in a metallic phase, the low temperature
 $P(M)$ peak gets pushed towards $M=0$ as the moments 
 collapse throughout the system. With increasing $T$, the distribution 
 broadens while the mean shifts
towards larger $M$ values.

\subsection{The zero temperature transition}
\label{sec:landau}

\begin{figure}[b]
\centering
\includegraphics[width=8.5cm,height=5.0cm]{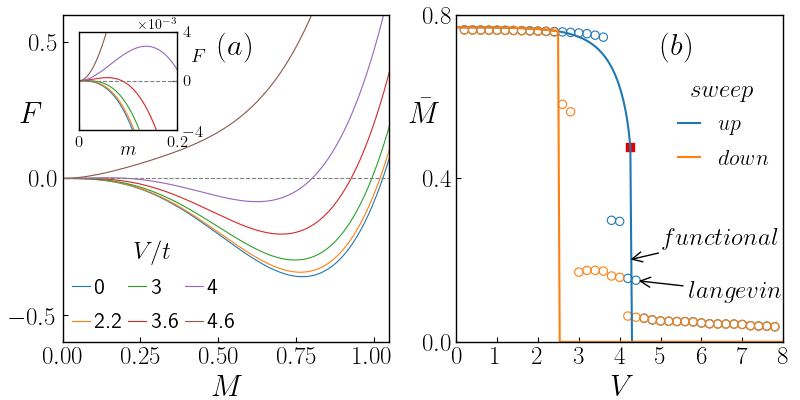}
\caption{(a) Effective functional for different values
  of $V/t$. For $V < V_{c}^{-}$ it has a unique minimum at
  large $M$. For $V_{c}^{-} < V \leq V_{c}^{-}$ it develops two
  minima (inset). For $V > V_{c}^{+}$ there is a unique minimum
  at $M = 0$. (b) The resulting moment profile which gets a sweep
  dependence in the coexistence region due to the presence of
  two minima. The open symbols are actual data points for $T=0.001t$.
  The red square indicates the point at which the moment profile
  jumps in the upward sweep according to the effective functional.
  In the effective functional. The large $V$ moment has been
  approximated to be zero in the effective description.
}
\label{fig:landau}
\end{figure}

The $T=0$ transition can be modeled by extremising an
effective functional $F$ of the form:
\begin{eqnarray}
\label{eq:landau}
F(M, V) &=& \frac{a(V)}{2}M^{2} - \frac{b}{3} M^{3} + \frac{c}{4}M^{4}
\cr
a(V) &=& a_{0}\left(e^{-V^{*}/V} - e^{-V^{*}/V^{-}_{c}}\right) 
\end{eqnarray}
where $M$ is the magnitude of the local moment, assumed to be uniform
across the system.
$a_{0}, V^{*}, b, c$ are fitting parameters which take positive
values. $a_{0}$, $V^{*}$ and $b$ can be determined in terms of the
moment magnitude at $V = 0$ and $V = V_{c+}$, and $\Delta V_{c}$.
The peculiar form of $a(V)$ ensures that the
finite $M$ minimum of $F$ remains almost pinned at $M = M(V=0)$
till $V\lesssim V_c^+$ and changes sharply across $V_c^+$. 
The moment profile does not depend on the parameter $c$, which
just sets the overall scale of $F$. It can be fixed by fitting the
low $T$, $P(M)$ at $V = 0$. Here, we have assumed the moment
amplitude in the large $V$ state to vanish. 
We find that $V^*\approx 4.8t$. The $V$ dependence of
$F$ has been shown in Fig.\ref{fig:landau}(a) and the resulting 
moment profile has been shown in Fig.\ref{fig:landau}(b).

For $V < V_c^-$, $F$ has a unique minimum at finite $M$.
For $V_{c}^{-} < V \leq V_{c}^{+}$, $F$ develops another minimum 
at
$M = 0$. Beyond $V_{c}^{+}$ only the $M = 0$ minimum survives.
 For the
upward voltage sweep the system remains stuck in the finite $M$
 minimum
till $V_{c}^{+}$ and then switches to the $M = 0$ minimum 
discontinuously.
A similar discontinuous transition happens in the downward
 sweep, in which
the starting state corresponds to the $M=0$ minimum, which 
changes abruptly
at $V_{c}^{-}$. This models the low $T$ coexistence and 
hysteresis.
However it is too simplistic to capture the finite $T$ transition,
for which one must take angular fluctuations of $\{\vec{M}\}$
into account.

\subsection{Finite temperature magnetism}
\label{sec:T_N}

\begin{figure}[b]
\centerline{
\includegraphics[width=8.7cm,height=4.0cm]{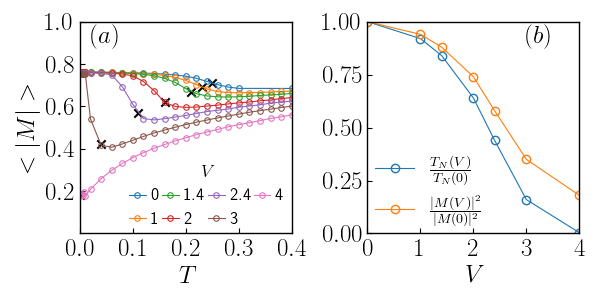}
}
\centerline{
\includegraphics[width=8.3cm,height=4.0cm]{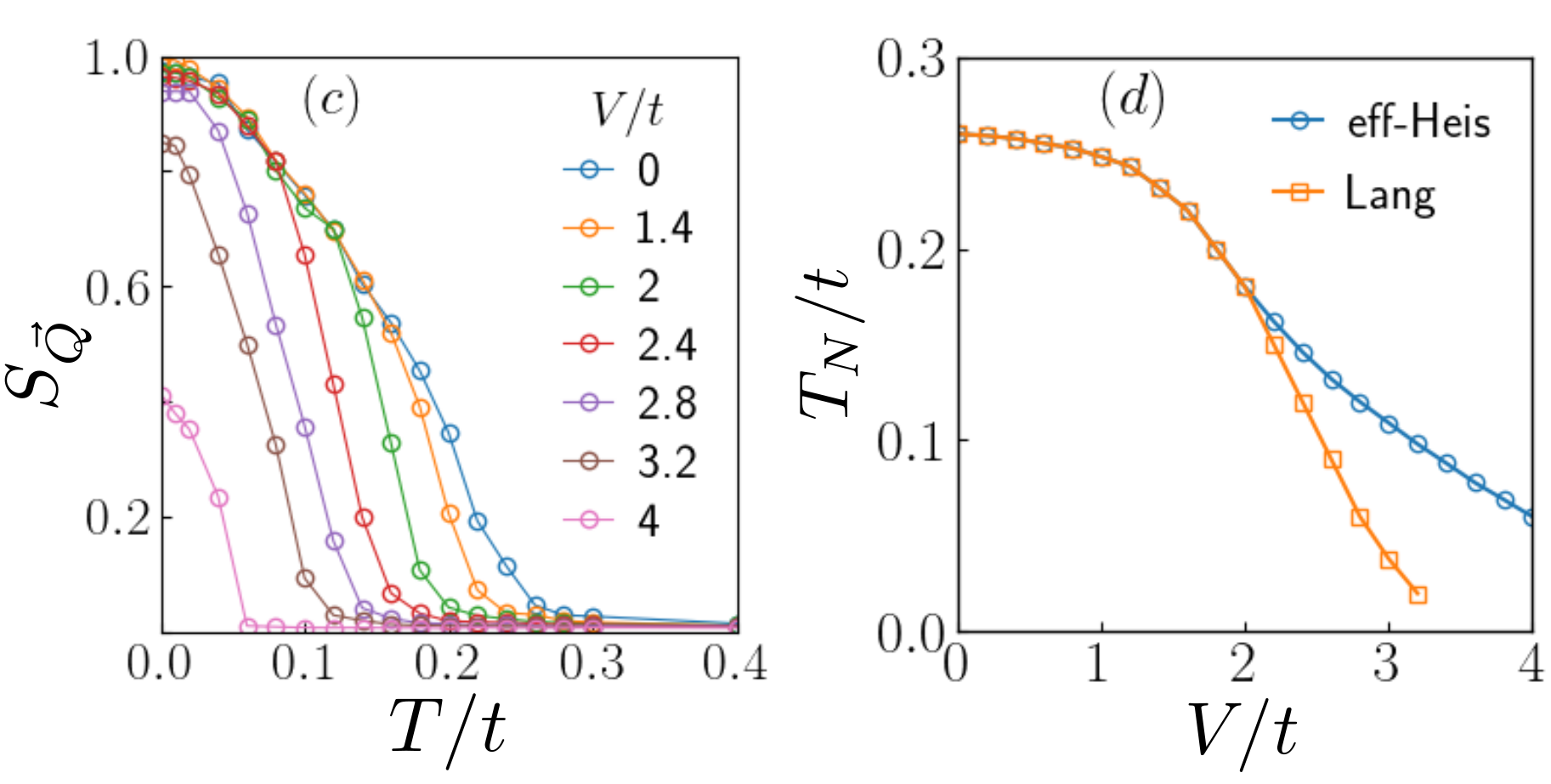}
}
\caption{(a)~Variation of average moment magnitude
${\bar M}$ with temperature for different values of bias voltage.
For each $V$ the corresponding N\'eel temperature ($T_{N}(V)$)
has been marked with a black cross on the trace.
${\bar M}(T)$ shows a minimum at a temperature $T_{*}(V)$.
(b)~Comparison of the $T_N(V)/T_N(0)$ with ${\bar 
M^2(V,T^*)}/{\bar M(0,0)}$ This suggests that a Heisenberg model 
with varying moment magnitude but $V, T$ indepedent coupling may 
describe the finite $V$ magnetism.
(c)~Temperature dependence of the magnetic structure
factor  at ${\vec Q} = (\pi,\pi,\pi)$ in the effective
Heisenberg model for different $V$, computed via
Monte Carlo.  (d)~Cmparison of the full Langevin based $T_N$ and 
that extracted from the effective Heisenberg model.
The correspondence works well upto intermediate $V$ and
breaks down in the metallic phase. }
\label{fig:effheis}
\end{figure}

As we have seen, 
the N\'eel temperature decreases with increasing $V$.
Making the crude assumption that the bare Heisenberg 
exchange scale,
$J = t^{2}/U$ at strong coupling, remains unchanged with $V$, 
we attempt to correlate the reduction in $T_{N}$
with the behaviour of the average moment magnitude
${\bar M}(V,T) =\langle  M \rangle$.
Fig.\ref{fig:effheis}(a) shows 
the variation of $ {\bar M}(V,T)$ with increasing 
$T$, for different $V$ values. We find that 
${\bar M}(V,T)$ behaves nonmonotonicaly
with temperature for $1 < V/t < 4$, 
and develops a minimum at
a temperature $T_{*}(V) \sim T_{N}(V)$. In 
Fig.\ref{fig:effheis}(b) we have compared
$\frac{T_{N}(V)}{T_{N}(0)}$ with
$\frac{{\bar M}^2(V,T^*)}{{\bar M}^2(0,0)}$ 
and find that they follow a similar trend with 
increasing $V$.

This correspondence 
suggests that an effective Heisenberg model for the
local moments may be able to describe the underlying physics.
To make progress we assume that the amplitude distributions
 of the local 
moments are same across the system, given by $P(M, V, T)$ 
and shown
in Fig.\ref{fig:pm}. In the insulating phase, increasing $T$ 
leads
to reduction in the local moment amplitudes, which is 
negligible at 
low $V$ but becomes significant near $V\lesssim V_{c}$ 
(above the
coexistence region). In the metal, however, the thermal 
broadening
of $P(M)$ leads to an increase in  mean moment size
with $T$. A combination of these two effects leads to the 
nonmonotonic $T$ depedence at intermediare $V$.

\begin{figure}[t]
\centering
\includegraphics[width=6.0cm, height=5.0cm]{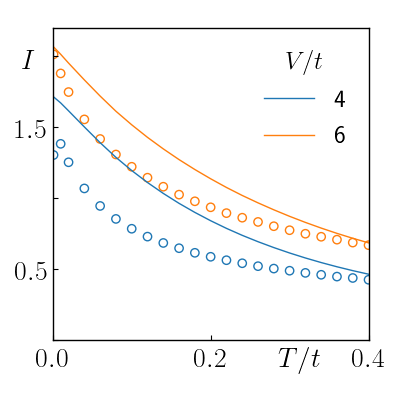}
\caption{Comparison of the approximate current (solid lines) 
with the exact result (open circles) in the paramagnetic metal 
phase for $V/t = 4, 6$.}
\label{fig:curr}
\end{figure}
Motivated by this, we attempt to explain the drastic 
reduction
in $T_{N}$ beyond $V \sim 0.5 V_{c}^{+}$ by invoking 
an effective
Heisenberg model, in which the moment magnitudes 
are determined
by the $P(M)$ distribution shown earlier. 
The effective Heisenberg model is then given by
$$
H_{eff} = \frac{t^{2}}{U} \sum_{ij}^{NN}
{\vec M}_i.{\vec M}_j
$$ 
where the magnitude of $\vec{M}_{i}$ at a site,
at a given $V$ and $T$, is obtained by sampling $P(M, V, T)$.
This approach, building in ``amplitude fluctuation'' of
the moments, gives a reasonable match with the
full Langevin calculation, for the structure factor
Fig.\ref{fig:effheis}(c) as well as the N\'eel temperature 
Fig.\ref{fig:effheis}(d), except very near $V_{c}$.

Rather than use the computed $P(M,V,T)$ we had tried
to use $P \propto e^{- F(M,V)/T}$ as amplitude weight.
That approach did not work, suggesting that the finite
temperature $F$ has a non trivial $T$ dependence.

The discussion above about moment magnitude and its
magnetic order pertains to the insulating state. At large
$V$ the low $T$ state is a metal, with essentially zero
magnetic moment. However there are thermally induced
moments in the metal, and their fluctuation serves
as a source of scattering. The next section provides
an analytic basis for this effect.

\begin{figure*}
\centering
\includegraphics[width=14cm,height=4cm]{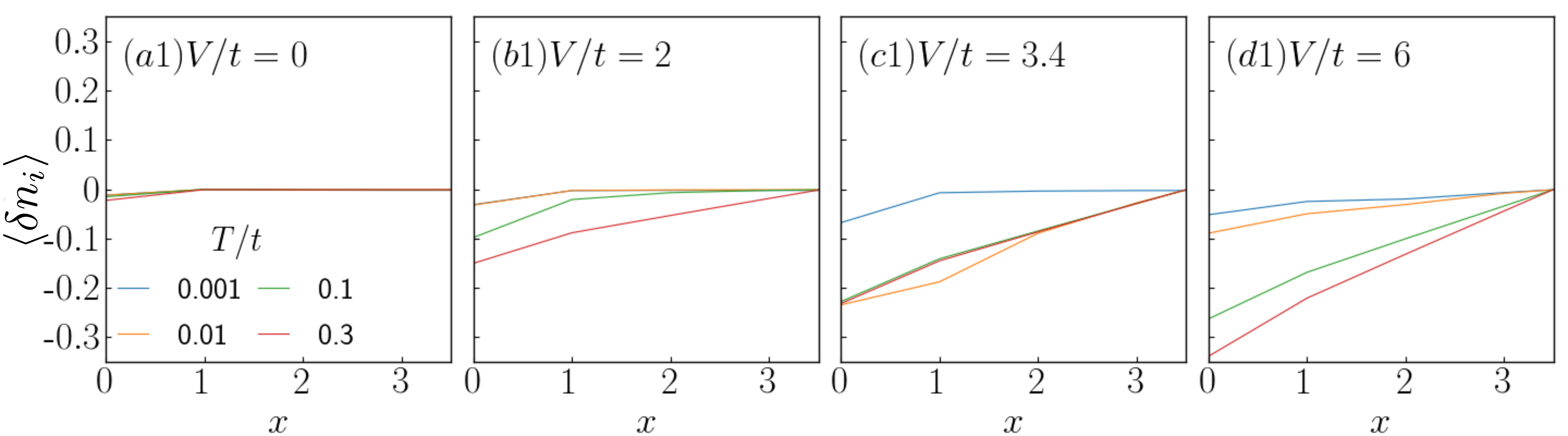}
\includegraphics[width=14cm,height=4cm]{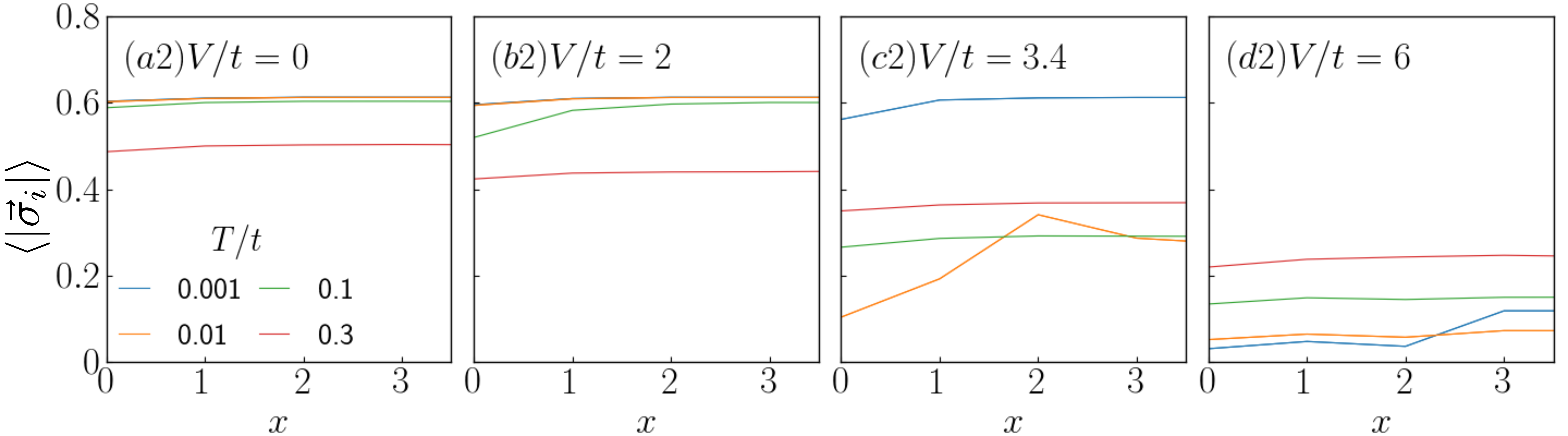}
\caption{(a1-d1) Variation of the average charge profile,
  along the longitudinal direction, with increasing temperature
  for $V/t = 0, 2, 3.4$ and $6$. With sufficient averaging the
  profile becomes antisymmetric about the center of the
  system,
  hence only the left half has been shown. (a2-d2)
  Variation of the
  average local moment magnitude along the
  longitudinal direction.
  The averaging leads to a symmetric profile
   across the center of the
  system.
}

\label{fig:aux}
\end{figure*}
\subsection{The bias stabilised metal}
\label{sec:curr}

One can set up an approximate calculation for the current
at large $V$, where the mean moment size gets quenched.
One can approximate the lesser Green's function
$\mathcal{G}^{<}(\tau,\omega)$, which enters the expression
for the current in Eq.4 in the main text, by setting up
a perturbation theory about the tight-binding limit.
\begin{subequations}
\label{eq:curr}
\begin{equation}
 \label{eq:Gless}
 \mathcal{G}^{<} = \frac{1}{2}\left(\mathcal{G}^{K}
   -\left(\mathcal{G}^{R}-\mathcal{G}^{A}\right)\right)
 \end{equation}
 \begin{equation}
 \begin{pmatrix}\mathcal{G}^{R} & \mathcal{G}^{K}\\
    0 & \mathcal{G}^{A}\end{pmatrix} =
 \begin{pmatrix}\left[g^{-1}\right]^{R}-\Sigma^{R} & 
\left[g^{-1}\right]^{K}-\Sigma^{K}\\
  0 & \left[g^{-1}\right]^{A}-\Sigma^{A}\end{pmatrix}^{-1}
\end{equation}
with,
\begin{equation}
 \label{eq:self_energy}
 \Sigma^{R,A,K}_{i,j;\alpha,\beta}(\tau,\omega) =
 U^2 M^{\alpha}_{i}(\tau)M^{\beta}_{j}(\tau)g^{R,A,K}_{ij}(\omega)
\end{equation}
\end{subequations}
where $g^{R,A,K}$ are the Green's functions of the 
connected tight-binding
system. The mean current is computed by averaging 
over the time-series
of $M$. This can be simplified further if one averages 
over the self-energy
instead of the Green's functions, assuming the 
distribution for $M$'s 
to be normal,
i.e.,
$\langle M^{\alpha}_{i}(\tau)M^{\beta}_{j}(\tau^{\prime})
\rangle\approx T\delta_{ij}\delta_{\alpha\beta}
\delta(\tau-\tau^{\prime})$.
So the averaged self-energy
\begin{equation}
 \label{eq:se_approx}
 \langle\Sigma^{R,A,K}_{i,j;\alpha,\beta}(\omega)\rangle =
 U^2 T\delta_{ij}\delta_{\alpha\beta}g^{R,A,K}_{ij}(\omega)  
\end{equation}
can be used to approximate the mean current. 
Fig.\ref{fig:curr} compares the
temperature dependence of the approximate current 
with the actual result
in the P-M phase. They seem to compare well for 
sufficiently large $V$,
given the drastic nature of the approximations.
This suggests that the current in the metallic phase
is essentially given by the tight-binding result for $V/t \geq 6$
as $T\rightarrow 0$.
For a finite system this has a finite value, and scales linearly
with the number of conduction channels which is proportional
to the cross sectional area $A$. Hence the resistance in the
metallic phase is finite even as $T\rightarrow 0$ within a finite
sized calculation, as is evident in Fig.\ref{fig:curr}, and would 
vanish only if $A\rightarrow\infty$. As the temperature increases,
thermal fluctuations of the background moments leads to
enhanced scattering which depletes the current further.

\subsection{Spatial variation of charge and magnetic moment}
\label{dens}

In Fig.\ref{fig:aux}, we show the  charge deviation (from $n_i = 1$)
and moment magnitude, defined as,
\begin{subequations}
  \begin{align}
    \langle\delta n_{i}\rangle &= 1 -
     \sum\limits_{i_{y},i_{z}}^{L_{y},L_{z}}
    \int\limits_{0}^{\tau_{M}}\frac{\mathrm{d}\tau}{\tau_{M}}\,
    \int\limits_{-D}^{D}
    \frac{\mathrm{d}\omega}{2\pi} \,
    \mbox{Tr}\left[\mathcal{
    G}^{K}_{ii}(\tau,\omega)\right]\\
    \langle |\vec{\sigma}_{i}|\rangle &= 
    \sum\limits_{i_{y},i_{z}}^{L_{y},L_{z}}
    \int\limits_{0}^{\tau_{M}}\frac{\mathrm{d}\tau}{\tau_{M}}\,
     \Biggl|\int\limits_{-D}^{D}
    \frac{\mathrm{d}\omega}{2\pi} \,
    \mbox{Tr}\left[\mathcal{G}^{K}_{ii}(\tau,\omega)
    \vec{\sigma}_P\right]\Biggr|
  \end{align}
\end{subequations}
where the trace is over the $2\times 2$ local spin subspace.
 At $V/t = 0$,
the charge deviation from $n_i = 1$ is vanishingly, as required,
at all temperature, while the moment magnitude
falls as the system is heated beyond $T_{N}$.
For $0 < V/t \leq 2.2$, the charge deviation is small and
concentrated at
the edge at very low $T$, and becomes linear at high $T$.
 The moment magnitude
also shows deviations at the edges, and gets diminished
throughout the system
with increasing $T$. For $2.2 < V/t < 3.6$, the charge
shows edge
deviations in low $T$ insulating phase, but becomes
 linear as one heats up the
system to reach the P-M phase. The moment magnitude
shows nonmonotonic
behaviour with temperature. For $V/t > 3.6$, the system
 remains in the P-M phase
at all $T$.
The charge profile remains linear, whose slope increases
with increasing $T$,
while the moment profile remains fairly flat and with
magnitude increasing with $T$.

\subsection{Size dependence}

The Langevin
scheme presented here leads to a numerically intensive
computation primarily
because of the presence of leads in the nonequilibrium
problem, and is
worsened by the presence of multiplicative noise. As a
result, for each site
in every time step we need to diagonalise the electronic
Hamiltonian twice.
Moreover, one needs to have a sufficiently long run
length in order to
achieve a steady state. All these considerations
constrain us to a
modest size, $8\times 4\times 4$.

However, to our benefit, we discovered that the voltage
driven
insulator-metal transition is strongly first order
for the 3D system. This
means the transition should be realised even in larger
systems
although the coexistence region may pick up a size
 dependence.
Furthermore, the magnetic transition at equilibrium is
easily captured
within our working size and several other studies
 have used even
smaller systems to study it. For the voltage driven
 problem, a larger
size may provide more resolution around the low
temperature
insulator-metal transition region of the phase diagram,
but it should
not change the essential features highlighted in this study.
We have also checked the size independence
of results for a few parameter points on a
$10\times 4\times 4$
system.

\subsection{Dimension dependence}

\begin{figure}[b]
\includegraphics[width=6.5cm,height=6.0cm]{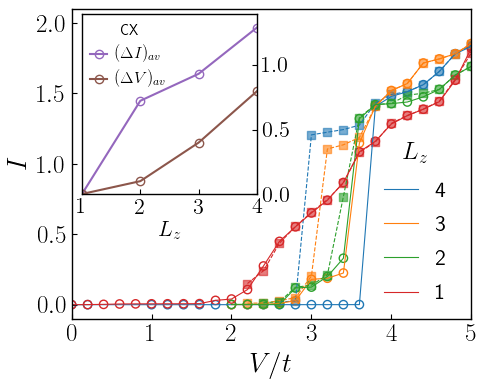}
\caption{Layer dependence of the $I$-$V$ characteristics at
$T/t = 0.001$. The size of the discontinuity and width of the
coexistence region progressively reduce with decreasing
thickness and vanish for a single layered 2D system.
The behaviour of average height
$(\Delta I)_{av}$ and average width $(\Delta V)_{av}$ of the
coexistence region with changing thickness has been shown in
the inset.}\label{supfig8}
\end{figure}

It is important to highlight the nature of the transition 
as we go from 
a 3D bar geometry to a 2D rectangular geometry by 
removing one
layer at a time, as shown in Fig.\ref{supfig8}. We 
find that the 
low temperature transition becomes progressively 
less abrupt as 
we reduce the number of layers, and for a single 
layer 2D system 
it becomes a crossover, which was studied in detail
 in Ref.\cite{neqmf}.
The hysteresis region systematically shrinks with reducing 
number of layers, and
for the single layer system no hysteresis is found. 
The ``softening''
of the jump in $I$-$V$ with reducing thickness of the 
sample has also 
been reported in an experiment\cite{cro-2019}.

\subsection{Connection with experiments}

Finally, the relevance of our results to real Mott
materials. Experimentally, the $I$-$V$ characterstics 
have a generic form across the 
transition metal oxides (TMO), e.g. vanadium oxides
\cite{vo2-2011,vo2-2015}, ruthenates \cite{cro-2013,cro-2019}, 
magnetites \cite{magnetite-2007,
magnetite-2009}, and some organics \cite{sabeth}. All these show 
a first order transition at low $T$ which gets weaker with increasing
temperature.
This aspect is well captured by our theory, unlike other
microscopic approaches. An experiment \cite{cro-2019} has also
reported that the jump in the $I-V$ becomes less abrupt
upon reducing sample thickness, which is also captured within
our scheme (Fig. \ref{supfig8}). An experiment on a 
multiorbital ruthenate
has reported suppression of N\'eel temperature with
 increasing
current \cite{cro-2019-II}. Some TMOs also undergo 
a temperature driven
structural transition at equilibrium. However, the transport 
measurements
have been made below this equilibrium transition 
temperature. Our theory
suggests that the transport characteristics can be 
explained via a purely
electronic mechanism. 

\section{Conclusions}

We have been able to construct a real time finite
temperature scheme to approach nonequilibrium effects in a
strongly correlated system. This Langevin equation approach 
simplifies the underlying Keldysh field theory by assuming
adiabaticity, {\it i.e}, electrons are much faster than 
magnetic degrees of freedom, and a thermal noise. With these
assumptions we could implement a numerical study of a Mott
insulator in a finite 3D geometry. We established a 
voltage sweep driven hysteretic insulator-metal transition 
at low temperature, the collapse of the Neel and pseudogap
temperature with increasing bias, and a thermally induced
insulator-metal transition at finite bias. In our analysis
the primary driver of the finite temperature effects is 
strong amplitude fluctuation of the local moments in the
bias induced first order landscape. This Langevin approach 
would
open up other nonequilibrium problems that have remained
inaccessible. 

\begin{acknowledgments}
We acknowledge use of the HPC clusters at HRI.
\end{acknowledgments}

\appendix\label{appendixA}
\section{Accessing equilibrium dynamics}
\label{sec:fp}

The Langevin scheme yields a time series for the auxiliary field 
$\vec{M}_{i}(\tau)$ starting from an initial configuration which may
be arbitrary. Thus the scheme allows for thermalisation of the
system to the equilibrium state. However, it remains to be 
ascertained 
that the system reaches the \emph{correct} equilibrium state, which
is nontrivial, given the \emph{multiplicative} nature of the noise. 
Moreover, the dissipation coefficient $\gamma_{i}(\tau)$ and the
diffusion $D_{i}(\tau)$ not only vary over sites, 
but also depend on the instantaneous configuration of the 
auxiliary fields.
Nevertheless, we shall show that even with such nontrivial 
parameters, 
the Langevin scheme converges to the \emph{correct} long 
time
equilibrium state. To do this we write the Fokker-Planck 
equation for the distribution function of the moments\cite{dudarev}
$P(\{\vec{m}_{i}(\tau)\},\tau) = \langle\ \prod_{i,a}
\delta(m^{a}_{i}-M^{a}_{i})\rangle$. 
\begin{subequations}\label{eq:fp}
	\begin{align}
	\frac{\partial\,P}{\partial\,\tau} &= 
	-\frac{\partial}{\partial\,m^{a}_{i}}
	\cdot\Biggl\{\biggl[\epsilon_{a b c}m^{b}_{i}F^{c}_{i}
	+ \alpha m^{a}_{i}m^{b}_{i}\left(F^{b}_{i}-T\tilde{D}_{i}
	\frac{\partial}{\partial\,m^{b}_{j}}\right)\nonumber\\
	&+\frac{1}{\alpha}\left(F^{a}_{i}-T\tilde{D}_{i}
	\frac{\partial}{\partial\,m^{a}_{i}}
	- \left(\delta_{ab} + m^{a}_{i}m^{b}_{i}\right)T
	\frac{\partial\,\tilde{D}_{i}}{\partial\,m^{b}_{i}}
	\right)\biggr] P\Biggr\}
	\tag{\ref{eq:fp}}
	\end{align}
\end{subequations}
	where the repeated indices are to be summed over, and
$ \tilde{D}_{i} = \frac{D_{i}\gamma_{i}}{2U} $
is the effective diffusion coefficient, and 
$
	\vec{F}_{i} = \langle\vec{\sigma}_{i}\rangle_{\{\vec{m}\}} -
	 \vec{m}_{i}
$ is a generalised force.
At steady state $\frac{\partial\,P}{\partial\,\tau} = 0$ leading 
to conservation
of probability current. For the dynamics at equilibrium, 
the following 
conditions are additionally met:
\begin{itemize}
	\item The force is \emph{conservative}, and hence, 
	can be derived 
	from the spin-fermion Hamiltonian as 
	$\vec{F}_{i} = -\frac{\partial \mathcal{H}}{\partial\vec{m}_{i}}$, 
	where,
	\begin{equation}
	\mathcal{H} = -t\sum_{<ij>,\sigma}
	\,d^{\dagger}_{i\sigma}d_{j\sigma}
	- U\sum_{i}\left(\vec{m}_{i}\cdot\vec{\sigma}^P_{i} +
	 |\vec{m}_{i}|^{2}\right)
	\end{equation}
	is the same Hamiltonian which was introduced in 
	Eq. 3g in the 
	Supplement\cite{supp}, with
	the scaling $\vec{M}^{c,q}\rightarrow U\vec{M}^{c,q}$ and
	including the classical stiffness of the auxiliary moments. 
	The charge
	field is fixed to the half-filling saddle point value. 
	\item The distribution is given by a Boltzmann form,
	\begin{equation}
		P(\{\vec{m}_{i}(\tau)\}) \propto 
		\mbox{Tr}_{el}\left(e^{-\beta\mathcal{H}} \right)
	\end{equation}
	Using this form in Eq. \ref{eq:fp}, we find that the 
	distribution becomes
	stationary if we assume $\tilde{D}_{i} = 1$
		which leads to,
		\begin{equation}
			D_{i} = \frac{2U}{\gamma_{i}}
			= \frac{\alpha}{1+\alpha^{2}|\vec{m}_{i}(\tau)|^{2}}
		\end{equation}
	This determines the diffusion coefficient in terms of the 
	known parameters 
	and the instantaneous background configuration.
\end{itemize}
Hence, we find that with a suitable choice of the dissipation and
diffusion coefficients, the Langevin scheme indeed leads to the 
\emph{correct} equilibrium state.

\section{Benchmarks}\label{appendixB}

\label{sec:bench}
 
\begin{figure}[b]
\centering
\includegraphics[width=4.2cm,height=4.4cm]{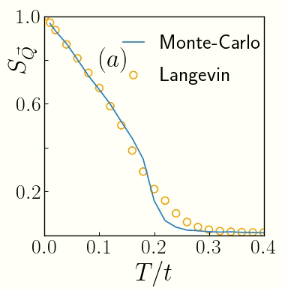}
\includegraphics[width=4.2cm,height=4.2cm]{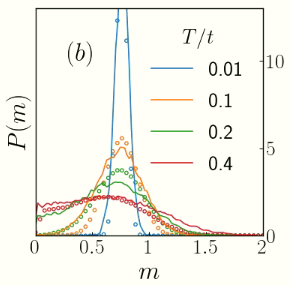}
\caption{(a) Comparison of temperature dependence of the structure
factor peak with equilibrium `classical' Monte carlo. The N\'eel
temperature $T_{N} = 0.28t$ at $U = 6t$. The two
curves coincide, except very close to $T_{N}$ .
(b) Comparison of moment distribution
for different temperatures. The distributions match at low and high
temperature and deviate slightly for low moment values near $T_{N}$.}
\label{fig:sfac_cmp}
\end{figure}

We benchmark our formulation against the classical Monte Carlo (MC)
formulation at equilibrium\cite{anamitra}. We compare the temperature
dependence of the structure factor (Eq.3 in the main text) peak across
the two formulations, in Fig.\ref{fig:sfac_cmp}(a). The two curves 
coincide for almost the entire range of temperature.
The magnetic transition temperature within Quantum Monte Carlo
(QMC) at $U/t = 6$ in 3D is about $0.3t$\cite{qmc}.
\par
We also compare the moment distribution defined in eq. \ref{eq:Pm}.
Fig.\ref{fig:sfac_cmp}(b) shows the comparison at different temperatures. 
The distributions match at low and high temperature, but slightly deviate 
for low $m$ close to the transition temperature.

%
%
\end{document}


\author{Arijit Dutta} 
\affiliation{Harish-Chandra Research Institiute,
HBNI, Chhatnag Road, Jhunsi, Prayagraj (Allahabad) 211019, India}
\affiliation{Institut f\"ur Theoretische Physik, 
Goethe-Universit\"at, 60438 Frankfurt am Main, Germany}

\author{Pinaki Majumdar}
\affiliation{Harish-Chandra Research Institiute,
HBNI, Chhatnag Road, Jhunsi, Prayagraj (Allahabad) 211019, India}

\title{Supplementary Material on ``The nonequilibrium thermal state
of a voltage biased Mott insulator''} 

\maketitle

\section{Real time path Integral}

Assuming the leads were connected far in the past, one can write the 
steady-state action for the system by discretizing the Keldysh contour 
shown in Fig.\ref{supfig1}. The generating functional is given by:
\begin{subequations}\label{orig_Z:main}
    \begin{equation}
    Z = \int\mathcal{D}\{\bar{c},c;\bar{d},d\} \, 
e^{i S\left[\bar{c},c;\bar{d},d\right]}\tag{\ref{orig_Z:main}}
    \end{equation}
    where $\left(\bar{c},c\right)$ and $\left(\bar{d},d\right)$ 
are the Grassmann fields for the lead and system fermions 
respectively. $S\left[\bar{c},c;\bar{d},d\right]$ is the complex 
time Keldysh action defined on the contour.
\begin{figure}[b]
\vspace{.5cm}
\includegraphics[width=\linewidth]{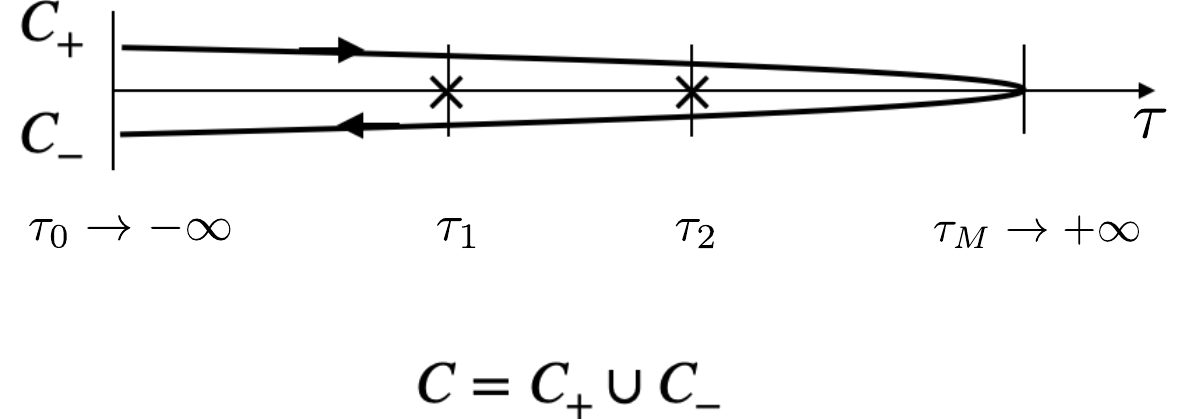}
\caption{The complex time Keldysh contour for nonequilibrium steady states. The
ends of the contour corresponding to the initial time $\tau_{0}$ and the maximum 
accessible time $\tau_{max}$ is taken to $\pm\infty$. In order to calculate 
observables,e.g., the two-point function, one makes insertions at intermediate 
times $\tau_1$ and $\tau_2$.}\label{supfig1}
\end{figure}
%
\begin{align}
S &= \int\limits^{\infty}_{-\infty}\mathrm{d}\tau\,
\left[\mathcal{L}_{sys}(\tau) + \mathcal{L}_{bath}(\tau) + \mathcal{L}_{coup}(\tau)
\right]\label{orig_Z:a}\\
\mathcal{L}_{sys}(\tau) &= \sum_{\substack{<ij>\\\sigma,s}}s \, 
\bar{d}^{s}_{i\sigma}(\tau)\left(i \partial_{\tau}+t\right)
  d^{s}_{j\sigma}(\tau)\nonumber\\
  &-U\sum\limits_{i,s}s \, n^{s}_{i\uparrow}(\tau)
n^{s}_{i\downarrow}(\tau)\label{orig_Z:b}\\
\mathcal{L}_{bath}(\tau) &= \sum_{\nu,\beta,s}s \, 
\bar{c}^{s}_{\nu\beta}(\tau)\left(i \partial_{\tau}-\epsilon_{\nu}\right)
c^{s}_{\beta}(\tau)\label{orig_Z:c}\\
\mathcal{L}_{coup}(t) &= \sum\limits_{\substack{<ij>\\\sigma,s}}v_{ij}
\left(\bar{c}^{s L}_{i\sigma}(\tau)d^{s}_{j\sigma}(\tau) + 
\bar{c}^{s R}_{i\sigma}(\tau)d^{s}_{j\sigma}(\tau) + g.c.\right)\label{orig_Z:d}
\end{align}
\end{subequations}
where $i$,$j$ are the lattice indices, $\sigma$ is the spin index, 
$\beta$ labels the leads and $s$ labels the contour. $s$ = $\pm$1 
for the upper and the lower contour fields respectively. At each
time slice, for every site we can rewrite the interaction term 
as:
\begin{subequations}\label{HS:main}
\begin{equation}
U n^{s}_{i\uparrow}n^{s}_{\downarrow} = \frac{U}{4}
\left(n^{s}_{i}\right)^{2} - U\left(\vec{\sigma}^{s}_{i}\cdot
\hat{\Omega}^{s}_{i}\right)^{2}\tag{\ref{HS:main}}
\end{equation}
where we have suppressed the time label for brevity. Here, 
$n^{s}_{i} = \bar{d}^{s}_{i\uparrow}d^{s}_{i\uparrow} +
\bar{d}^{s}_{i\downarrow}d^{s}_{i\downarrow}$ 
is the local density, $\vec{\sigma}^{s}_{i} = \frac{1}{2}\sum
\limits_{\alpha\beta}\bar{d}^{s}_{i\alpha}
\vec{\sigma}^{P}_{\alpha\beta}d^{s}_{i\beta}$ is the local spin 
operator and $\hat{\Omega}^{s}_{i}$
is an arbitrary SO(3) vector. $\vec{\sigma}^P$ is the $2\times 2$
Pauli vector.

Each of the two terms can be decomposed by a Hubbard-Stratonovich 
(HS) transformation. The first transformation introduces a 
``charge field'' $\phi_{i}(\tau)$:
\begin{equation}
e^{-i\frac{sU}{4}\left(n^{s}_{i}(\tau)\right)^{2}} \propto \int\mathrm{d}
\phi^{s}_{i}(\tau) \, e^{i \left(\frac{s}{U}(\phi^{s}_{i}(\tau))^2 
- s\phi^{s}_{i}(\tau)n^{s}_{i}(\tau)\right)}\label{HS:charge}
\end{equation}

While the HS transformation on the second term can be written in terms 
of an O(3) ``spin field'' $\vec{M}_{i}(\tau)$:

\begin{align}
&e^{isU\left(\vec{\sigma}^{s}_{i}(\tau)\cdot\hat{\Omega}^{s}_{i}(\tau)\right)^{2}} 
\propto\nonumber\\
&\int\mathrm{d}^{3}\vec{M}^{s}_{i}(\tau)\,e^{i\left(-\frac{s}{U}
(|\vec{M}^{s}_{i}(\tau)|)^2+ s\vec{M}^{s}_{i}(\tau)\cdot\vec{\sigma}^{s}_{i}(\tau)
\right)}\label{HS:spin}
\end{align}
\end{subequations}
Upon introducing the auxiliary fields the action becomes quadratic in
the fermions, which can be integrated out to get the following
effective action:
\begin{subequations}\label{S_diss}
\begin{align}
\tilde{S}\left[\phi,M\right] &= -\iota Tr\ln\left[\iota\check{G}^{-1}
                                (\tau,\tau^{\prime})\right]\nonumber\\
  &+ \frac{1}{U}\int\mathrm{d}t\sum_{i}
\left[\phi^{c}_{i}(\tau)\phi^{q}_{i}(\tau) - \vec{M}^{c}_{i}(\tau)\cdot
\vec{M}^{q}_{i}(\tau)\right]\tag{\ref{S_diss}}
\end{align}
with
\begin{equation}\label{Ginv}
\check{G}^{-1}\left(t,t^{\prime}\right) = 
\begin{pmatrix}
\hat{G}^{-1}_{R} \, & \,
\hat{G}^{-1}_{12}\\\\
\hat{G}^{-1}_{21} \, & \,
\hat{G}^{-1}_{A}
\end{pmatrix}
\end{equation}
where the components of $\mathbb{G}^{-1}$ in the 2 $\times$ 2 Keldysh space are given by:
\begin{align}
\hat{G}^{-1}_{R}(\tau,\tau^{\prime}) &= \left(\iota\partial_{\tau} - \hat{\mathcal{H}}^{c}(\tau)\right)
\delta(\tau-\tau^{\prime})+ \hat{\Gamma}^{R}(\tau,\tau^{\prime})\label{Ginv:R}\\
\hat{G}^{-1}_{A}(\tau,\tau^{\prime}) &= \left(\iota\partial_{\tau} - \hat{\mathcal{H}}^{c}(t)\right)
\delta(\tau-\tau^{\prime}) + \hat{\Gamma}^{A}(\tau,\tau^{\prime})\label{Ginv:A}\\
\hat{G}^{-1}_{K}(\tau,\tau^{\prime}) &= \hat{\Gamma}^{K}_{ij\alpha}(\tau,\tau^{\prime})
\delta_{\alpha\alpha^{\prime}}\nonumber\\
&+ \left(\hat{G}^{-1}_{R}\circ F - F\circ\hat{G}^{-1}_{A}\right)(\tau,\tau^{\prime})
\label{Ginv:K}\\
\left[\hat{G}^{-1}_{12}(\tau,\tau^{\prime})\right]_{\substack{ij;\\
\alpha\alpha^{\prime}}} &= \frac{1}{2}\left(\vec{M}^{q}_{i}(\tau)\cdot
\vec{\sigma}^P_{\alpha\alpha^{\prime}} - \phi^{q}_{i}(\tau)\delta_{\alpha\alpha^{\prime}}
\right)\nonumber\\
&\delta_{ij}\delta(\tau-\tau^{\prime}) + 
\left[\hat{G}^{-1}_{K}(\tau,\tau^{\prime})\right]_{\substack{ij;\\\alpha\alpha^{\prime}}}\label{Ginv:12}\\
\left[\hat{G}^{-1}_{21}(\tau,\tau^{\prime})\right]_{\substack{ij;\\
\alpha\alpha^{\prime}}} &= \frac{1}{2}\left(\vec{M}^{q}_{i}(\tau)\cdot
\vec{\sigma}^P_{\alpha\alpha^{\prime}} - \phi^{q}_{i}(\tau)\delta_{\alpha\alpha^{\prime}}
\right)\nonumber\\
						&\delta_{ij}\delta(\tau-\tau^{\prime})\label{Ginv:21}\\
\hat{\mathcal{H}}^{c}_{ij}(\tau) =& \left(\phi^{c}_{i}(\tau)\mathbb{1}_{2}-
\vec{M}^{c}_{i}(\tau)\cdot\vec{\sigma}^P\right)\delta_{ij}
- t_{<ij>}\mathbb{1}_{2}\label{H_c}
\end{align}
\end{subequations}
where $\mathbb{1}_{2}$ is the $2\times 2$ identity matrix.
$\hat{\mathcal{H}}^{c}(\tau)$ is a time-dependent
Hamiltonian which depends on the `classical' component of the
auxiliary fields. $\hat{\Gamma}^{R,A,K}(\tau,\tau^{\prime})$ are dissipation terms
which enter the action as a result of integrating out the leads. $F(\tau,\tau^{\prime})$
is the distribution function of the disconnected system, and $\circ$ denotes
convolution. The `classical' and `quantum' components of the
auxiliary fields are linear combinations of the fields introduced
in the H-S transformations in Eqs.\ref{HS:charge} and \ref{HS:spin}.

\begin{align}
    \vec{M}^{c} = \frac{1}{2}\left(\vec{M}^{+} + \vec{M}^{-}\right),\,\,\,\,
    \vec{M}^{q} = \left(\vec{M}^{+} - \vec{M}^{-}\right)
\end{align}
where we have suppressed the time and other labels for notational 
brevity. A similar transformation holds for the $\phi$ fields.
We have mapped the original fermionic action for the dissipative 
Hubbard model into an action containing time dependent auxiliary 
fields. The mapping is formally exact up to this point.

\section{Approximations}

Next, we derive an effective thermal fluctuation theory by making
a series of approximations: (i) We fix the charge field ($\phi$) to it's
classical saddle point at equilibrium, i.e., $\phi^{c}_{i}, \phi^{q}_{i} = 0$.
This fixes the overall density to half-filling, but allows local density
fluctuations due to dynamics of the spin-field ($\vec{M}$).
(ii) We perform a cumulant expansion of the action to second order
in $\{M^{q}\}$ fields, introduce a `noise' by decoupling the
quadratic term, and evaluate the `classical' saddle point to
obtain a stochastic equation of motion (EOM) for the $\{M^{c}\}$. 
(iii) We simplify the
EOM by performing a semiclassical expansion of the two-point functions
to obtain a `Langevin' equation in terms of the `slow' time coordinate.
The noise kernel is assumed to be Gaussian, which can be justified
in the high temperature limit.

\subsection{Cumulant expansion}

The $\check{G}^{-1}$ introduced in Eq.\ref{Ginv} can be decomposed into
a Green's function $\check{G}^{-1}_{c}$, which depends only on the `classical'
field and a self-energy $\check{\Sigma}_{q}$, which depends only on the
`quantum' field.
\begin{align}
  \check{G}^{-1} &= \left(\check{\mathbb{1}} + \check{\Sigma}_{q}\circ\check{G}_{c}
  \right)\circ\check{G}^{-1}_{c}\nonumber\\
\check{\Sigma}_{q}(\tau,\tau^{\prime}) &= \frac{1}{2}\vec{M}^{q}_{i}(\tau)\cdot
\vec{\sigma}^P_{\alpha\alpha^{\prime}}\delta_{ij}\delta(\tau-\tau^{\prime})\otimes\sigma^{x}_{K}
\end{align}
where $\sigma^{x}_{K}$ denotes the structure in $2\times2$ Keldysh space.
We expand the action in Eq.\ref{S_diss} to second order in $\check{\Sigma}_{q}$.
\begin{subequations}\label{S_cum}
  \begin{align}
    \tilde{S} = \tilde{S}^{0}+\tilde{S}^{1}+\tilde{S}^{2} + ...\tag{\ref{S_cum}}
  \end{align}
  with
\begin{equation}
		\tilde{S}^{0} = -\iota \mbox{Tr}\ln\left[\iota\check{G}^{-1}_{c}(\tau,\tau^{\prime})\right] = 0\\
\end{equation}
$\tilde{S}^{0}$ vanishes due to the causality relation between the retarded and
advanced Green's functions.

\begin{align}
\tilde{S}^{1} &= \sum_{i}\int\mathrm{d}\tau\left(\mbox{Im}\left[\mbox{Tr}
\left(\hat{G}^{K}_{ii}(\tau,\tau)\vec{\sigma}\right)\right]-\frac{\vec{M}^{c}_{i}(\tau)}{U}\right)
\cdot\vec{M}^{q}_{i}(\tau)\\
\tilde{S}^{2} &= \frac{\iota}{2}\int\mathrm{d}\tau\int\mathrm{d}\tau^{\prime}
\sum_{ij;ab}M^{q}_{ia}(\tau)\left[\hat{\Pi}^{K}(\tau,\tau^{\prime})\right]^{ab}_{ij}M^{q}_{jb}(\tau^{\prime})
\end{align}
where
\begin{align}
  \label{eq:sigmaK}
&\left[\hat{\Pi}^{K}(\tau,\tau^{\prime})\right]^{ab}_{ij}\equiv
\mbox{Tr}\Biggl[\left(\hat{G}^{K}_{ij}(\tau,\tau^{\prime})\sigma^{a}\hat{G}^{K}_{ji}(\tau^{\prime},\tau)
\sigma^{b}\right)\nonumber\\
&+\left(\hat{G}^{R}_{ij}(\tau,\tau^{\prime})\sigma^{a}\hat{G}^{A}_{ji}(\tau^{\prime},\tau)\sigma^{b}\right)
+\left(\hat{G}^{A}_{ij}(\tau,\tau^{\prime})\sigma^{a}\hat{G}^{R}_{ji}(\tau^{\prime},\tau)\sigma^{b}\right)\Biggr]
\end{align}
Now, we decompose the term quadratic in $M^{q}$.
\begin{align}
  e^{\iota\tilde{S}^{2}} &= e^{-\frac{1}{2}M^{q}\circ\hat{\Pi}^{K}\circ M^{q}}\nonumber\\
  &\propto \int\mathcal{D}[\xi]\, \exp\left(-\frac{1}{2}\xi\circ
  \left[\hat{\Pi}^{K}\right]^{-1}
  \circ\xi - \iota\, \xi\circ M^{q}\right)
\end{align}
\end{subequations}
Hence, this adds a term to the coefficient of $M^{q}$ in $\tilde{S}^{1}$ and
additionally, the generating functional is reweighted by the quadratic piece
in $\xi$.

We obtain the equation of motion by requiring that the first order variation
w.r.t the `q' fields must vanish at $\phi^{q}=0, \vec{M}^{q}=0$. This gives us the
following equation for the `c' fields:
\begin{align}
\mbox{Tr}\left[\hat{G}^{K}_{ii}(\tau,\tau)\vec{\sigma}^P\right]
&=\vec{M}^{c}_{i}(\tau) - \vec{\xi}_{i}(\tau)\label{eom:M}\nonumber\\
\langle\xi^{a}_{i}(\tau)\xi^{b}_{j}(\tau^{\prime})\rangle &= \left[\hat{\Pi}^{K}(\tau,\tau^{\prime})\right]^{ab}_{ij}
\end{align}

\subsection{Semiclassical expansion}

Transforming to Wigner coordinates allows us to write.
\begin{align}
  \hat{G}^{K}(\tau,\tau) = 
\hat{G}^{K}(\tau,\tau_{r}=0) = \int\mathrm{d}\omega\, \hat{G}^{K}(\tau,\omega)
  \label{wig}
\end{align}
In what follows, we would write a series expansion for $\hat{G}^{K}(\tau,\omega)$ in
powers of $\hbar$. For this purpose we first construct a series expansion for
$\hat{G}^{R}(\tau,\omega)$.

\subsubsection{The retarded Green's function $\hat{G}^{R}$}

The retarded Green's function obeys the Dyson's equation
\begin{subequations}\label{GR}
  \begin{equation}
   (\hat{G}^{-1\,R} \circ \hat{G}^{R})(\tau_1,\tau_2) = \delta(\tau_1-\tau_2)
   \hat{\mathbb{1}}\tag{\ref{GR}}
  \end{equation}
  Transforming to Wigner coordinates allows us to expand the LHS in a Kramers-Moyal
  series\cite{kamenev},
  \begin{align}
  &\left(\omega-\hat{\mathcal{H}}^{c}(\tau)+\hat{\Gamma}^{R}(\omega)\right)
      \hat{G}^{R}(\tau,\omega)\nonumber\\
    &+ \frac{\iota\hbar}{2}\Biggl(\partial_{\tau}\hat{\mathcal{H}}^{c}(\tau)\partial_{\omega}\hat{G}^{R}
    -(\hat{\mathbb{1}}+\partial_{\omega}\hat{\Gamma}^{R})\partial_{\tau}\hat{G}^{R}\Biggr)
     +O(\hbar^{2}) = \hat{\mathbb{1}}\label{GR:a}
  \end{align}
  where $\tau \equiv \frac{\tau_1 + \tau_2}{2}$ is the center-of-mass time, and $\Omega$ is the
  fourier conjugate to the relative time $\tau_r \equiv \tau_1 - \tau_2$. Furthermore, we have assumed
  that $\hat{\Gamma}^{R}$ depends only on the relative time. This is true
  at steady state. 
  
  This expansion relies on the condition that $\hbar\ll$ (timescale 
  for $\vec{M}$ fluctuations)$\times$(energy-scale for electronic
  excitations). The timescale for magnetic fluctuations is set by
  $J^{-1}$, where $J\sim\frac{t^2}{U}$ is the magnetic exchange
  scale, while the energy-scale for electronic excitations is set
  by the electronic bandwidth given by $U$, for $U\ll t$. Setting
  $\hbar = 1$, we then have the condition $J\ll U$ for 
  the expansion to be valid.
  
  Inverting the above Eq. gives us
  \begin{align}
    &\hat{G}^{R}(\tau,\omega) = \hat{\mathcal{G}}^{R}(\tau,\omega)\nonumber\\
    &- \frac{\iota\hbar}{2}
    \Biggl(\hat{\mathcal{G}}^{R}\partial_{\tau}\hat{\mathcal{H}}^{c}(\tau)\partial_{\omega}
                             \hat{\mathcal{G}}^{R}
                             + \hat{\mathcal{G}}^{R}(\hat{\mathbb{1}} + \partial_{\omega}
    \hat{\Gamma}^{R})\partial_{\tau}\hat{\mathcal{G}}^{R}\Biggr) + O(\hbar^{2})\label{GR:b}
  \end{align}
  where,
  \begin{align}
    \hat{\mathcal{G}}^{R}(\tau,\omega) = \left(\omega\hat{\mathbb{1}} - \hat{\mathcal{H}}^{c}(\tau)
    +\hat{\Gamma}^{R}(\omega)\right)^{-1}\label{GR:c}
  \end{align}
is the ``adiabatic'' retarded Green's function which depends only 
on the instantaneous configuration of the background field, 
and not on its history.

Now, for any matrix $\hat{A}(\alpha)$, we have $\partial_{\alpha}\hat{A} =
-\hat{A}\left(\partial_{\alpha}\hat{A}^{-1}\right)\hat{A}$. Using this we can rewrite
Eq.\ref{GR:b} to $O(\hbar)$ as
\begin{align}
  \hat{G}^{R}(\tau,\omega) = \left(\hat{\mathbb{1}} + \frac{\iota\hbar}{2}\left[
  \hat{\mathcal{G}}^{R}\partial_{\tau}\hat{\mathcal{H}}^{c}\, \bm{,} \,
  \hat{\mathcal{G}}^{R}\left(\hat{\mathbb{1}}+\partial_{\omega}
  \hat{\Gamma}^{R}\right)\right]\right)\hat{\mathcal{G}}^{R}(\tau,\omega)
\end{align}

where $\left[\, \bm{,} \,\right]$ denotes the commutator bracket.
\end{subequations}

\subsubsection{The Keldysh Green's function $\hat{G}^{K}$}

Knowing $\hat{G}^{R}$ and $\hat{G}^{A}$ to any order in $\hbar$ allows one
to construct the $\hat{G}^{K}$ to that order provided the distribution function 
is known apriori. In our case, the distribution function in the disconnected 
system $F(\Omega) = tanh(\omega/2T)$ gets corrections due
to hybridisation with the leads, as is apparent from the form of 
$\hat{G}^{-1}_{K}(\tau,\tau^{\prime})$ defined in Eq.\ref{Ginv:K}.
From the structure of $\check{G}^{-1}$,
it follows that
\begin{subequations}\label{GK}
  \begin{equation}
    \hat{G}^{K}(\tau_1,\tau_2) = -\hat{G}^{R}(\tau_1,\tau^{\prime})\circ
    \hat{G}^{-1}_{K}(\tau^{\prime},\tau^{\prime\prime})\circ\hat{G}^{A}(\tau^{\prime\prime},\tau_{2})\tag{\ref{GK}}
  \end{equation}
  Transforming to Wigner coordinates and implementing the Kramers-Moyal expansion,
  as above, allows us to write
  \begin{align}
    &\hat{G}^{K}(\tau,\omega) = \hat{\mathcal{G}}^{K}(\tau,\omega)\nonumber\\
      &+ \frac{\iota\hbar}{2}\left[\hat{\mathcal{G}}^{R}\partial_{\tau}\hat{\mathcal{H}}^{c}
    \, \bm{,} \, \hat{\mathcal{G}}^{R}\left(\hat{\mathbb{1}} + \partial_{\omega}
    \hat{\Gamma}^{R}\right)\right]F\hat{\mathcal{G}}^{R} - H.c. + \left(..\right)\label{GK:semi}
\end{align}
where,
\begin{align}
    \hat{\mathcal{G}}^{K}(\tau,\omega) &= F(\omega)\left(\hat{\mathcal{G}}^{R}(\tau,\omega)-\hat{\mathcal{G}}^{A}(\tau,\omega)\right)\nonumber\\
    &- \hat{\mathcal{G}}^{R}(\tau,\omega)\hat{\Gamma}^{K}(\omega)\hat{\mathcal{G}}^{A}(\tau,\omega)\label{GK:adia}
  \end{align}
\end{subequations}
The terms denoted by $\left(..\right)$ in Eq.\ref{GK:semi} can be dropped as their 
contribution is negligible for a gapped system.
Now, we can write the matrix elements of adiabatic Green's function in the $2\times 2$
spin subspace as:
\begin{subequations}
\begin{align}
  \hat{\mathcal{G}}^{R}_{ij}(\tau,\omega) = g^{R}_{ij}(\tau,\omega)\hat{\mathbb{1}}_{2\times 2}
  + \vec{h}^{R}_{ij}(\tau,\omega)\cdot\vec{\sigma}
\end{align}
and similarly, for,
\begin{align}
  \hat{\chi}^{R}_{ij}(\tau,\omega) \equiv -\iota\hbar\left[\hat{\mathcal{G}}\partial_{\tau}
  \hat{\mathcal{H}}^{c},\hat{\mathcal{G}}^{R}\left(\hat{\mathbb{I}}
  +\partial_{\omega}\hat{\Gamma}^{R}\right)\right]
\end{align}
we can write,
\begin{align}
  \hat{\chi}^{R}_{ij}(\tau,\omega) &= \rho^{R}_{ij}(\tau,\omega)\hat{\mathbb{I}}_{2\times 2}
  + \vec{\lambda}^{R}_{ij}(\tau,\omega)\cdot\vec{\sigma}
\end{align}

Plugging these in eq.\ref{GK:semi} we obtain
\begin{align}
&\mbox{Tr}\left[\hat{G}^{K}_{ii}(\tau,\omega)\sigma^{a}\right] = \mbox{Tr}\left[\hat{
\mathcal{G}}^{K}_{ii}(\tau,\omega)\sigma^{a}\right]+\sum_{j}\Biggl(-\Gamma_{ij}(\tau,\omega)
\partial_{\tau}M^{a}_{j}\nonumber\\
&+\left(\vec{\Lambda}_{ij}(\tau,\omega)\times\partial_{\tau}\vec{M}_{j}\right)_{a}
+\vec{\beta}^{a}_{ij}(\tau,\omega)\cdot\partial_{\tau}\vec{M}_{j}\Biggr)\label{GK:full}
\end{align}
where
\begin{align}
  \Gamma_{ij} &= \left(\rho^{R}_{ij}g^{R}_{ji}-\vec{\lambda}^{R}_{ij}
  \cdot\vec{h}^{R}_{ji} + R\rightarrow A\right)\\
  \vec{\Lambda}_{ij} &= \left(\rho^{R}_{ij}\vec{h}^{R}_{ji}
  -\vec{\lambda}^{R}_{ij}g^{R}_{ji} + R\rightarrow A\right)\label{eq:alpha}\\
  \vec{\beta}^{a}_{ij} &= \left(\vec{\lambda}^{R}_{ij}h^{R}_{a,ji}
  -\lambda^{R}_{a,ij}\cdot\vec{h}^{R}_{ji} + R\rightarrow A\right)
\end{align}
\end{subequations}
Finally, from eq.\ref{wig} we know that the equal time Keldysh Green's
function which enters the EOM eq'\ref{eom:M} is obtained by performing
an integral over $\omega$ of eq.\ref{GK:full}.
\begin{subequations}
  \begin{equation}
  \int\mbox{Im}\left[\mbox{Tr}\left(\hat{\mathcal{G}}^{K}_{ii}
  (\tau,\omega)\vec{\sigma}^P\right)\right]
  \mathrm{d}\omega \equiv \langle\vec{\sigma}_{i}(\tau)\rangle_{\{\vec{M}\}}
  \end{equation}
  where $\langle\vec{\sigma}_{i}(\tau)\rangle$ is the instantaneous expectation value of
the electron spin obtained under the adiabatic approximation. 

Evaluating the coefficients $\Gamma_{ij}$, $\vec{\Lambda}_{ij}$ and $\vec{\beta}^{a}_{ij}$
for an arbitrary $\{\vec{M}\}$ configuration is very hard. We simplify them 
by performing a strong coupling expansion and retaining just the leading order terms. 
$\vec{\beta}^a$ drops out in this process. Further, we postulate the following form
for the remaining coefficients,
\begin{align}
  \int\mathrm{d}\omega\,\Gamma_{ij}(\tau,\omega) &\approx \frac{1}{\gamma_{i}(\tau)}\delta_{ij}\\
  \int\mathrm{d}\omega\,\vec{\Lambda}_{ij}(\tau,\omega)
  &\approx \frac{\alpha}{\gamma_{i}(\tau)}\vec{M}_{i}(\tau)\delta_{ij}\\
  \left[\Pi^{K}(\tau,\tau^{\prime})\right]^{ab}_{ij} &\approx 2D_{i}(\tau)T
  \delta_{ij}\delta_{ab}\delta(\tau-\tau^{\prime})
\end{align}
\end{subequations}
where $\gamma_{i}(\tau)$, $D_{i}(\tau)$ and $\alpha$ are unknown parameters,
which we shall determine subsequently,
and $T$ is the temperature. 
$\alpha$ is the dimensionless 
\emph{Gilbert damping}\cite{gilbert,lakshmanan} 
parameter which provides a relaxational torque to the angular 
degrees of freedom, $\gamma_{i}$ contributes to longitudinal 
damping and $D_{i}$ are position dependent diffusion coefficients.
The uncorrelated form of the noise kernel can be motivated by expanding 
$\Pi^{K}(\tau,\Omega)$ about the homogeneous antiferromagnetic
state in powers of $\Omega/\Delta$ and $t/\Delta$, where $\Omega$ is the 
characterisic scale of two particle excitation ($J\sim t^2/U$) and $\Delta$ is the 
gap in the single particle spectrum, and taking the limit $\Omega\ll T$.  
The noise vanishes as $T\rightarrow 0$ as 
a result of this approximation, but the actual noise survives even at zero 
temperature in a quantum system.
In Appendix A, we show $D_{i}(\tau) = 2U/\gamma_{i}(\tau)$ for \emph{detailed balance} to be 
satisfied at equilibrium. Hence, one obtains the form of the evolution 
equation for $\{M\}$, as given in the main text.
 
Next shall fix $\gamma$. For this we
cast the Langevin equation into a
Landau-Lifshitz-Gilbert(LLG) form by solving for
$\frac{\partial \vec{M}}{d\tau}$,
\begin{align}\label{eq:llg}
  &\frac{d\vec{M}_{i}}{d\tau} = 
  -A_{i}(\tau)\vec{M}_{i}\times\left(\langle\vec{\sigma}_{i}\rangle
  +\vec{\xi}_{i}\right)\nonumber\\
  &+ B_{i}(\tau)\vec{M}_{i}\times\left(\vec{M}_{i}\times
  \left(\langle\vec{\sigma}_{i}\rangle+\vec{\xi}_{i}
  \right)\right) +
  \gamma_{i}(\tau)\left(\vec{M}_{i}
  -\langle\vec{\sigma}_{i}\rangle+\vec{\xi}_{i}\right)
\end{align}
with,
$A_{i}(\tau)\equiv\frac{\gamma_{i}(\tau)\alpha}{1+\alpha^{2}|M_{i}(\tau)|^{2}}$
and 
$B_{i}(\tau)\equiv\frac{\gamma_{i}(\tau)\alpha^{2}}{1+\alpha^{2}|M_{i}(\tau)|^{2}}$.
The first term is the `Bloch' term which leads to precessional motion 
of the spins. In order to capture the correct spin 
wave dispersion $A_{i} = 2U$ must be satisfied, which follows from the
subleading term in the strong coupling expansion of 
$\langle\vec{\sigma}_{i}\rangle$. This means that the time dependence
in $A_{i}(t)$ must drop out, which leads us to the condition,
\begin{equation}\label{eq:gamma}
	\gamma_{i}(\tau) = \frac{2U}{\alpha}\left(1+\alpha^{2}|\vec{M}_{i}(\tau)|^{2}\right)
\end{equation}
and hence, $B_{i} = 2U\alpha$ also becomes independent of time.

$\alpha$ remains the only free parameter in the Langevin scheme.
It is dimensionless, and the static properties do not depend sensitively
on $\alpha$. It can be estimated by 
evaluating Eq.\ref{eq:alpha} for a two site system. A tedious, but 
straightforward, calculation suggests $\alpha = (U/t)^{2}$.
Hence, all the parameters which were introduced `by-hand' get 
fixed upon further considerations and the final formulation 
does not have any free parameters. Having said that, we must
note that unlike the other two parameters the value of $\alpha$
doesn't get fixed by any consistency condition, but rather through
a calculation. We shall test the sanctity of this result by 
benchmarking our scheme against equilibrium Monte-Carlo 
results.

An alternative, and perhaps more intuitive approach, would be to show 
that the Langevin equation for the auxiliary 
fields maps to the well studied Landau-Lifshitz-Gilbert (LLG) equation at 
strong coupling. This is easy to see once we expand the electronic spin 
in powers of $t/U$ in terms of the instantaneous background configuration,
\begin{equation}
	\langle\vec{\sigma}_{i}(t)\rangle 
	= \vec{M}_{i}(\tau) -  \frac{J}{2U}\sum_{j\in NN}\vec{M}_{j}(\tau)
\end{equation}
where $J = 4t^{2}/U$. Substituting this 
in Eq. \ref{eq:llg} and projecting the dynamics on the spheres defined 
by $|\vec{M}_{i}| = 1/2$ we get\cite{dudarev},
\begin{equation}\label{eq:effLLG3}
	\frac{d\vec{M}_{i}}{d\tau} = \vec{M}_{i}
	\times\left(\vec{B}_{i}
	+\vec{\xi}_{i}\right)
	- \tilde{\alpha}\vec{M}_{i}\times
	\left(\vec{M}_{i}\times\left(\vec{B}_{i}
	+\vec{\xi}_{i}\right)\right)
\end{equation}
where $\vec{B}_{i} = J\sum_{i+z\in NN}\vec{M}_{i+z}$ and
$\tilde{\alpha} = \frac{3\alpha}{4}-\frac{1}{\alpha}$.
The configuration dependence of the parameters drops out.
This is the celebrated LLG equation.

Thus, we find that the Langevin formulation reproduces the known
limits consistently. However, in the context of Hubbard model the 
observable quantities are electronic correlations which can be 
calculated within the \emph{adiabatic} approximation by solving 
the electronic problem in the instantaneous
auxiliary field background. Thus, this gives us a time series for 
electronic operators as well, from which we can compute various 
static and dynamic correlators.